\documentclass[journal,transmag]{IEEEtran}
\usepackage{cite}
\usepackage{graphicx}
\usepackage{float}
\usepackage{url}
\usepackage{amsmath,amsfonts,amssymb}
\usepackage{booktabs}
\usepackage{caption}
\usepackage{subcaption}
\usepackage[export]{adjustbox}
\usepackage[update,prepend]{epstopdf}
\usepackage[lined,algonl,ruled]{algorithm2e}
\usepackage{balance}
\usepackage{acronym}
\usepackage{color}
\usepackage{hyperref}
\hypersetup{
 bookmarksopen=true,
 bookmarksnumbered=true,
 colorlinks=true,
 citecolor=blue,
 linkcolor=red,
 linktocpage=true,
 linkbordercolor={1 1 1},
 hypertexnames=false,
 plainpages=true,
 pdfsubject={},
 pdfkeywords={},
 pdftitle={A Hybrid Approach for Speech Enhancement Using MoG Model and Neural Network Phoneme Classifier},
 pdfauthor={Shlomo E. Chazan, Jacob Goldberger and Sharon Gannot}
}

\newcommand{\X}{\mathbf{X}}

\newcommand{\Y}{\mathbf{Y}}
\newcommand{\Z}{\mathbf{Z}}
\newcommand{\x}{\mathbf{x}}
\newcommand{\y}{\mathbf{y}}
\newcommand{\z}{\mathbf{z}}
\newcommand{\w}{\mathbf{w}}
\newcommand{\h}{\mathbf{h}}
\newcommand{\vv}{\mathbf{v}}
\newcommand{\comment}[1]{}
\newcommand{\T}{{\scriptscriptstyle \top}}

\ifCLASSINFOpdf

\else

\fi

\acrodef{PSD}{power spectral density}
\acrodef{NN}{neural network}
\acrodef{MFCC}{mel-frequency cepstral coefficients}
\acrodef{MMSE}{minimum mean square error}
\acrodef{ASR}{automatic speech recognition}
\acrodef{STSA}{short-time spectral amplitude estimator}
\acrodef{LSAE}{log spectral amplitude estimator}
\acrodef{OMLSA}{optimally modified log spectral amplitude}
\acrodef{IMCRA}{improved minima controlled recursive averaging}
\acrodef{STFT}{short-time Fourier transform}
\acrodef{DFT}{discrete Fourier transform}
\acrodef{MoG}{Mixture of Gaussians}
\acrodef{r.v.}{random variable}
\acrodef{p.d.f.}{probability density function}
\acrodef{NN}{neural network}
\acrodef{EM}{expectation-maximization}
\acrodef{SPP}{speech presence probability}
\acrodef{CMVN}{cepstral mean and variance normalization}
\acrodef{NN-MM}{neural network mixture-maximum}
\acrodef{PESQ}{perceptual evaluation of speech quality}
\acrodef{SNR}{signal to noise ratio}
\acrodef{DAE}{deep auto-encoder}
\acrodef{LLR}{log likelihood ratio}
\acrodef{WSS}{weighted spectral slope}
\acrodef{Covl}{overall quality}
\acrodef{Csig}{speech distortion}
\acrodef{Cbak}{background distortion}
\acrodef{WSJ}{wall street journal}
\acrodef{SVM}{support vector machine}
\acrodef{IBM}{ideal binary mask}
\acrodef{IRM}{ideal ratio mask}

\begin{document}

\title{A Hybrid Approach for Speech Enhancement Using MoG Model and Neural Network Phoneme Classifier}

\author{\IEEEauthorblockN{Shlomo E. Chazan,
Jacob Goldberger and
Sharon Gannot \IEEEmembership{Senior Member, IEEE,}}
\thanks{Shlomo E. Chazan, Jacob Goldberger and Sharon Gannot are with the Faculty of Engineering, Bar-Ilan University, Ramat-Gan, 5290002, Israel (e-mail: Shlomi.Chazan@biu.ac.il; Jacob.Goldberger@biu.ac.il; Sharon.Gannot@biu.ac.il).}\\%
\IEEEauthorblockA{Faculty of Engineering, Bar-Ilan University, Ramat-Gan, 5290002, Israel}}

\maketitle

\begin{abstract}
In this paper we present a single-microphone speech enhancement
algorithm. A hybrid approach is proposed merging the generative \ac{MoG} model and the discriminative \ac{NN}. The proposed algorithm is executed in  two phases, the training phase, which does not recur, and the test phase. First, the noise-free speech \ac{PSD} is modeled as a \ac{MoG}, representing the phoneme based diversity in the speech signal. An \ac{NN} is then  trained with phoneme labeled database for phoneme classification with \ac{MFCC} as the input features. Given the phoneme classification results, an \ac{SPP} is obtained using both the generative and discriminative models. Soft spectral subtraction is then executed while simultaneously, the noise estimation is updated.  The discriminative \ac{NN} maintain the continuity of the speech and the generative phoneme-based \ac{MoG} preserves the speech spectral structure.  Extensive experimental
study using real speech and noise signals is provided. We also compare the proposed algorithm
 with alternative speech enhancement algorithms. We show that we obtain a significant improvement
over previous methods in terms of both speech quality measures and speech recognition results.

\end{abstract}


%

\section{Introduction}

\IEEEPARstart{E}{nhancing} noisy speech received by a single microphone is a widely-explored problem. A plethora of approaches can be found in the literature~\cite{38}. Although many current devices are equipped with multiple microphones, there are still many applications for which only a single microphone is available.

One such application involves \ac{ASR} systems. It is well-known that such systems are sensitive to mismatch between the train and test environments. Enhancing the noisy speech signal prior to the application of the \ac{ASR} system, might alleviate the performance degradation caused by the environment. Nonstationary noise environments are usually more challenging, since the speech enhancement algorithm should adapt to the changing statistics of the additive noise.

The celebrated \ac{STSA} and \ac{LSAE}~\cite{20,39} are widely-used model-based algorithms. The \ac{OMLSA} estimator and  in particular the \ac{IMCRA} noise estimator are specifically tailored to nonstationary noise environments~\cite{17,32}. However, fast changes in noise statistics often yields the \emph{musical noise} phenomenon.

Recently, \ac{NN} techniques gained a lot of popularity due to theoretical and algorithmic progress, and  the availability of more data and more processing power. Unlike past learning algorithms for NN, it is now possible to infer the parameters of the \ac{NN} with many layers, and hence the name \emph{deep learning}. Deep learning methods were mainly applied to speech recognition and lately, for speech enhancement as well. \ac{NN} and a \ac{DAE} were used as a nonlinear filters in~\cite{2} and~\cite{25}, respectively. The networks are trained on stereo (noisy and clean) audio features, to infer the complex mapping from noisy to clean speech. An experimental study with this approach is shown in~\cite{47}.  The \ac{NN} reduces the noise level significantly, yet, the enhanced signals still suffer from noticeable \emph{speech distortion}.

Other methods  attempt to train an \ac{NN} to find a mask, which classifies the time-frequency bins into speech/noise classes. Given the binary mask, the noisy bins are decreased. In~\cite{1} for instance, a \ac{SVM} is used to estimate the \ac{IBM} for speech separation from non-speech background interference. An \ac{NN} is trained to find the input features for the \ac{SVM}. A simpler approach is to train the \ac{NN} itself to find the \ac{IBM}. Different targets for the \ac{NN} are presented in  ~\cite{45}. The \ac{IBM} has shown advantageous in terms of intelligibility~\cite{49}. Yet, the binary mask is known to introduce artifacts such as \emph{musical noises}. For intelligibility tasks, this might not be problematic, though for speech enhancement the \ac{IBM} is not sufficient. To circumvent  this phenomenon, in~\cite{46} the \ac{NN} is trained to find the \ac{IRM}, which is a soft mask. A comparison between the \ac{IBM} and the \ac{IRM} is presented in~\cite{44}. The soft mask is better  than the  binary mask in terms of speech quality.  These approaches do not use models nor assumptions for their speech enhancement. However, they are trained with specific noise types, resulting in poor enhancement in an untrained noise environment. To cope with this problem,   in~\cite{24} the \ac{NN}  was trained with more than 100 different types of noise. Nevertheless, in real-life  where the number of noise types are not limited, this approach may not be satisfactory.

Training-based algorithms, such as MixMax~\cite{3}, were also developed. These algorithms are performed in two phases, the training phase and the test phase. In the training phase the parameters of the model are found, usually with an unsupervised machine  learning algorithms, such as the \ac{EM} algorithm in~\cite{3}. In the test phase, the enhancement is carried out using the learned model parameters. One weakness of the algorithm is that the speech parameters are  found  in an unsupervised manner that ignores the phoneme-based  structure of speech. Another drawback of the MixMax algorithm is that the noise parameters are estimated once at the beginning of the utterance and then are kept  fixed  during the entire utterance. This enhancement approach    is not always sufficient for real-life noises.

In this paper, we apply a hybrid algorithm, which integrates the generative model-based approach with the discriminative  \ac{NN} tool. As in~\cite{3}, we use a two phase algorithm. In the training phase, the clean speech is modeled with a phoneme-based  \ac{MoG} that is built using phoneme labeled database. A \ac{NN} is then trained  to classify clean\footnote{The NN is trained on clean signals in order to remain general and  not to adjust the network for certain noise types.} time-frame features as one of the phonemes from the phoneme-based \ac{MoG}.  Once the training phase is over, the training does not recur.   With the \ac{NN} estimated phonemes, an \ac{SPP} is calculated in the test phase using the generative model. Soft spectral subtraction is then carried out  using the SPP, while, simultaneously, the  noise estimation is updated. The continuity of the speech is maintained using the \ac{NN} that uses context frames in addition to the current frame. In addition, the \ac{NN}  assists the calculation of the SPP. Furthermore, the phoneme-based \ac{MoG} and the soft \ac{SPP} preserve the spectral structure of the speech thus alleviating the musical noise phenomenon. This approach utilizes the benefits of both the generative and the discriminative methods to alleviate the drawbacks of the mentioned above algorithms.

The rest of the paper is organized as follows. In Section~\ref{sec:model}, a generative model is presented. Section~\ref{sec:algorithm}  presents the proposed enhancement algorithm and describes its implementation in details.
 A comprehensive experimental results using speech databases in various noise types are presented in Section~\ref{sec:results}. In Section~\ref{sec:analysis} the building blocks of the algorithm are  analyzed. Finally, some conclusions are drawn and the paper is summarized in Section~\ref{sec:summer}.

\section{A Generative Noisy Speech Model}\label{sec:model}

In this section, a generative model of the noisy speech signal is presented. We follow the model proposed by N{\'a}das et al.~\cite{6} that was utilized in~\cite{3}.

The  following  notation is used throughout the paper. Uppercase letters are used for random variables, lower case for a given value and a boldface symbols denotes vectors.

\subsection{Maximization approximation}

Let $x(t)$ and $y(t) \quad 0<t<T$ denote the speech and noise signals, respectively. The observed noisy signal $z(t)$ is given by
\begin{equation}
	z(t)=x(t)+y(t).
	\label{xyzt}
\end{equation}
Applying the \ac{STFT} with frame length set to $L$ samples and overlap between successive frames set to $3L/4$ samples to $z(t)$ yields $Z(n,k)$ with $n$ the frame index and $k=0,1,\ldots,L-1$ the frequency index. The frame index $n$ is henceforth omitted for brevity, whenever applicable.

Let $\Z$ denote the $L/2+1$ dimensional log-spectral vector, defined by
$${Z}_k=\log|Z(k)|,  \quad k=0,1,\ldots,L/2.$$
Note that the other frequency bins can be obtained by the symmetry of the \ac{DFT}. Similarly, we define $\X$ and $\Y$ to be the log-spectral vectors of the speech and noise signals, respectively.

It is  assumed that the noise is statistically independent of the speech signal. Furthermore, it is assumed that both the speech and noise are zero-mean stochastic processes.
  Due to these assumptions the following approximation can be justified:
 $$|Z(k)|^2\approx |X(k))|^2+|Y(k)|^2$$
 hence
 $$Z_k \approx \log(e^{X_k}+e^{Y_k}).$$

Following N{\'a}das et al.~\cite{6}, the noisy log-spectral can be further approximated:
 \begin{equation}
	 \Z\approx \max(\X,\Y)
	 \label{maximization}
 \end{equation}
 where the maximization is component-wise over the elements of  $\X$ and $\Y$. This approximation was found useful for speech recognition~\cite{6}, speech enhancement~\cite{3,23}  and speech separation tasks~\cite{26,29}.
In a speech enhancement task,  only the noisy signal $\Z$ is observed, and the aim is to estimate the clean speech $\X$.

\subsection{Clean speech model - Phoneme based \ac{MoG}}
It is well-known that a speech utterance can be described as a time-series of phonemes,  i.e. speech is uttered by pronouncing a series of phonemes~\cite{27}. In our approach, we give this observation a probabilistic description, namely the log-spectral vector of the clean speech signal, $\X$, is modelled by a \ac{MoG} distribution, where each mixture component is associated with a specific phoneme. Unlike~\cite{3}, that uses unsupervised clustering of the speech frames, we use here a supervised clustering, explicitly utilizing the labels of the phonemes of the training speech signals.
Based on the MoG model, the probability density function $f(\x)$ of the clean speech $\X$, can be written as
\begin{equation}
f(\x)=\sum_{i=1}^m c_if_i(\x)=\sum_{i=1}^m c_i\prod_{k}f_{i,k}(x_k)
\end{equation}
where $m$ is the number of mixture components and
\begin{equation}
f_{i,k}(x_k)=\frac{1}{\sqrt{2\pi}\sigma_{i,k}}\exp
\left\{-\frac{(x_k-\mu_{i,k})^2}{2\sigma^2_{i,k}}\right\}.
\label{nnMoG}
\end{equation}
Let $I$ be  the phoneme indicator \ac{r.v.} associated with the \ac{MoG} \ac{r.v.}  $\X$, i.e. $p(I=i)=c_i$. The term $f_{i}(\x)$ is the Gaussian \ac{p.d.f.} of $\X$ given that  $I=i$. The scalar $c_i$ is the probability of the $i$-th mixture and  $\mu_{i,k}$ and $\sigma_{i,k}$ are the mean and the standard deviation of the $k$-th entry of the $i$-th mixture Gaussian, respectively. Due to the Fourier transform properties, we neglect any residual correlation between the frequency bins. Since for each class $I=i$ the \ac{r.v.} $\X$ is Gaussian, the frequency bins are also statistically independent. Consequently, the covariance matrix of each mixture component is diagonal.
To set the MoG parameters we used the phoneme-labeled TIMIT database~\cite{11,18}  as described in Sec.~\ref{subsec:trainNN}.

\subsection{Noisy speech model}
Let $\Y$ define the log-spectral vector of the noise signal, and let $g(\y)$ denote the \ac{p.d.f.} of $\Y$.  As with the log-spectral vector of the speech signal, it is assumed that the components of $\Y$ are statistically independent. For simplicity, $g(\y)$ is modeled as a single Gaussian, with diagonal covariance i.e.,
\begin{equation}
g(\y)=\prod_{k}g_k(y_k)
\end{equation}
where
\begin{equation}
\begin{aligned}
g_{k}({y_k})=\frac{1}{\sqrt{2\pi}\sigma_{Y,k}}
\exp\left\{-\frac{(y_k-\mu_{Y,k})^2}{2\sigma^2_{Y,k}}\right\}.
\end{aligned}
\end{equation}
Initial estimation and adaptation the noise parameters will be explained in Sec.~\ref{subsec:noiseadapt}.

 Using the maximum assumption in the log-spectral vector of the noisy speech $\Z =\max(\X,\Y)$, as explained above, it can be verified~\cite{6} that the \ac{p.d.f.} of $\Z$ is given by the following mixture model:
 \begin{equation}
  h(\z)=\sum_{i=1}^m c_ih_i(\z)=\sum_{i=1}^m c_i\prod_{k}h_{i,k}(z_k)
  \label{hz}
 \end{equation}
 where
\begin{equation}
 h_{i,k}(z_k) =f_{i,k}(z_k)G_k(z_k)+F_{i,k}(z_k)g_k(z_k)
\end{equation}
such that $F_{i,k}(x)$ and $G_k(y)$ are the cumulative distribution functions of the Gaussian densities $f_{i,k}(x)$ and $g_k(y)$, respectively. The term $h_{i}(\z)$ is the \ac{p.d.f.} of $\Z$ given that $I=i$.

The generative modeling described above was nicknamed MixMax~\cite{3,6}, since it is based on the maximum assumption and on the modelling of the clean speech as a (Gaussian) mixture \ac{p.d.f.} and the noisy speech is modeled as the maximum of the clean speech and the noise signal. Originally, the mixture components were not associated with phonemes, but rather learned in an unsupervised manner.

\section{The Neural-Network MixMax Algorithm}\label{sec:algorithm}
In this section, we describe the proposed enhancement algorithm. In Sec.~\ref{subsec:mmse} we remind the \ac{MMSE} estimator based on the MixMax model~\cite{3,6}. We then propose in Sec.~\ref{subsec:soft} a new variant of the estimator that utilizes the same model but allows for better noise reduction. In Sec.~\ref{subsec:NN} an \ac{NN} approach is introduced as a tool for accurate phoneme classification.
Issues regarding the training of the \ac{NN} are discussed in Sec.~\ref{subsec:trainNN}.
 Finally, test-phase noise adaption is discussed in Sec.~\ref{subsec:noiseadapt}.

\subsection{The MMSE based approach}\label{subsec:mmse}
An \ac{MMSE} of the clean speech $\x$ from measurement $\z$ is obtained by the conditional expectation $\hat{\x}=E(\X|\Z=\z)$. Note, that since the \ac{p.d.f.} of both $\x$ and $\z$ is non-Gaussian, this estimator is not expected to be linear.
Utilizing the generative model described in the previous section we can obtain a closed-form solution for the MMSE estimator as follows.
\begin{equation}
\hat{\x}=\sum_{i=1}^m p(I=i|\Z=\z)  E(\X|\Z=\z,I=i).
\label{conditional expextation}
\end{equation}
The posterior probability  $p(I=i|\Z=\z)$ can be easily obtained from~\eqref{hz} by applying the Bayes' rule:
 \begin{equation}
  p(I=i|\Z=\z)=\frac{c_ih_i(\z)}{h(\z)}.
  \label{p_i_z}
 \end{equation}
Since the Gaussian covariance matrices of both the speech and the noise models are diagonal,
  we can separately compute
 $$\hat{\x}_i = E(\X|\Z=\z,I=i)$$
    for each frequency bin.
For the $k$-th frequency bin we obtain:
\begin{eqnarray}
\hat{x}_{i,k}& = & E(X_k|Z_k= z_k,I=i) \label{exjIi}\\
&=& \rho_{i,k} z_k + (1-\rho_{i,k}) E  (X_{k}|  X_{k}<z_k ,I=i) \nonumber
 \end{eqnarray}
such that
\begin{equation}
\rho_{i,k}=p(Y_k<X_{k}|Z_k=z_k,I=i)=\frac{f_{i,k}(z_k)G_k(z_k)}{h_{i,k}(z_k)}
\label{rhodef}
\end{equation}
and for the second term in~\eqref{exjIi}:
\begin{equation}
E(X_k| X_k<z_k, I=i)=\mu_{i,k}-\sigma^2_{i,k}\frac{f_{i,k}(z_k)}{F_{i,k}(z_k)}.
\label{exkzk}
\end{equation}
The closed-form expression for the MMSE estimator of the clean speech $\hat{\x}=E(\X|\Z=\z)$~\cite{6} is obtained from~\eqref{conditional expextation},\eqref{exjIi},\eqref{rhodef},\eqref{exkzk}. These expressions are the core of the MixMax speech enhancement algorithm proposed by Burshtein and Gannot~\cite{3}. In their approach the MoG parameters of the clean speech are inferred from a database of speech utterances utilizing the \ac{EM} in an unsupervised manner.

 \subsection{Soft mask estimation of the clean speech}\label{subsec:soft}
Assuming the maximization model in~\eqref{maximization} is valid, $\rho_{i,k}$ was obtained in~\eqref{rhodef}.
Summing over all the possible mixture components, we obtain:
\begin{equation}
\rho_k = \sum_{i=1}^m p(I=i|\Z=\z) \rho_{i,k} = p(X_k > Y_k| \Z=\z).
\label{rhok}
\end{equation}
The term $\rho_k$ can be interpreted as the probability that given the noisy speech vector $\z$, the $k$-th frequency bin of the current log-spectral vector $\z$ is originated from the clean speech and not from the noise. The probability $\rho_k$  can thus be viewed as a training-based \ac{SPP} detector, namely the probability that the designated time-frequency bin is dominated by speech. Consequently,  $(1-\rho_{k})$ can be interpreted as the posterior probability that the $k$-th bin is dominated by noise.

Using $\rho_k$ and~\eqref{conditional expextation},\eqref{exjIi} the $k$-th frequency bin of the MMSE estimator $\hat{\x}=E(\X|\Z=\z)$ can be recast as follows:
\begin{equation}
\Hat{x}_k = \rho_k z_k + (1-\rho_k) E(X_k|X_k<z_k,\Z=\z).
\label{hatxk}
\end{equation}
Hence, given the generative model, the enhancement procedure in~\eqref{hatxk}, substitutes the frequency bins identified as noise with the a priori value drawn from the \ac{MoG} model and using~\eqref{exkzk}.

The structure of voiced speech \ac{PSD} consists of dominant spectral lines which recur at multiples of the fundamental frequency (known as \emph{pitch}).  The \ac{PSD} of different speakers pronouncing the same phoneme share similar properties, but are never identical. Hence, the \ac{MoG} parameters inferred from multiple speakers, is never individualized to the current speaker and therefore cannot represent the specific periodicity. The phoneme-based \ac{MoG} parameters are only capable of preserving the general structure of an averaged phoneme.  This phenomenon might lead to \emph{residual noise} even when the algorithm identifies the noise correctly.

To circumvent this phenomenon, we propose to substitute the optimal estimator that uses the \ac{MoG} parameters with a simpler estimate based on the spectral substraction paradigm, namely:
 \begin{equation*}
E (X_k| X_k<z_k, \Z=\z)
 \end{equation*}
is substituted by:
 \begin{equation*}
 z_k-\beta
 \end{equation*}
 where $\beta$ is a noise reduction level. It is well-known that the basic spectral subtraction method is prone to \emph{musical noise}~\cite{30}\cite{31}. In our proposed method, the estimator also incorporates the soft mask deduced from the \ac{SPP}, thus potentially alleviating the musical noise phenomenon.

 Substituting  $(z_k-\beta)$ in~\eqref{hatxk} we obtain  the following simplified expression for the estimated clean speech:
 \begin{equation}
 \hat{x}_k = \rho_{k}\cdot z_k + (1-\rho_{k})\cdot({z_k}-\beta)
  \end{equation}
 or, equivalently
 \begin{equation}
 \hat{x}_k =  z_k - (1-\rho_{k})\cdot\beta
 \label{soft}
 \end{equation}
which can be interpreted as \ac{SPP}-driven (soft) spectral subtraction algorithm.

 \subsection{Neural network for phoneme classification} \label{subsec:NN}
The gist of our approach is the calculation of the \ac{SPP} $\rho_k$~\eqref{rhok}. This calculation necessitates two terms, $\rho_{i,k}$ which is given by~\eqref{rhodef} and the posterior phoneme probability $p_i\triangleq p(I=i | \Z=\z)$.
Utilizing the generative model defined in Section~\ref{sec:model}, $p_i$ is obtained from~\eqref{hz} by applying the Bayes' rule:
\begin{equation}
p_i=\frac{c_ih_i(\z)}{h(\z)}.
\label{p_i_z2}
\end{equation}
This approach exhibits some major shortcomings. Estimating the required noise statistics is a cumbersome task, especially in time-varying scenarios. Furthermore, as the calculation in~\eqref{p_i_z2} is carried out independently for each frame, continuous and smooth speech output cannot be guaranteed.

In our approach, (unlike~\cite{3}) we adopt a \emph{supervised} learning approach in which each mixture component of the clean speech is associated with a specific phoneme. Hence the computation of the mixture index
posterior probability   becomes a phoneme classification task (based on the noisy speech). To implement this supervised classification task, we substitute~\eqref{p_i_z2} with an \ac{NN} that is known to be significantly better than \ac{MoG} models for phoneme classification tasks (see e.g.~\cite{10}).

The \ac{NN} is trained on a phoneme-labeled clean speech. For each log-spectral vector, $\z$, we calculate the corresponding \ac{MFCC} features (and their respective delta and delta-delta features). To preserve the continuity of the speech signal, 9 MFCC vectors are concatenated (the current feature vector, 4 past vectors and 4 future vectors) to form the feature vector, denoted $\vv$, which is a standard feature set for phoneme classification. This feature vector is used as the input to the \ac{NN}, and the phoneme label as the corresponding target. The phoneme-classification \ac{NN} is trained on clean signals. However, as part of the speech enhancement procedure, we apply it to noisy signals.
To alleviate the mismatch problem between train and test conditions, we use a standard preprocessing stage for robust phoneme classification, namely \ac{CMVN}~\cite{12}.

The \ac{SPP} $\rho_k$ is calculated using~\eqref{rhok}, which requires both $\rho_{i,k}$ and $p_i$.
While $\rho_{i,k}$ is calculated from the generative model using~\eqref{rhodef}, we propose to replace~\eqref{p_i_z2} for calculating $p_i$ by a better phoneme-classification method.

It is therefore proposed, to infer the posterior phoneme probability by utilizing the \emph{discriminative} \ac{NN}, rather than resorting to the generative MoG model:
\begin{equation}
p_i^{\textrm{NN}}=p(I=i|\vv;\textrm{NN}).
\label{p_i_z_NN}
\end{equation}
Note, that the compound feature vector $\vv$ is used instead of the original log-spectrum $\z$.
Finally, the SPP $\rho_k$ is obtained using~\eqref{rhodef} and~\eqref{p_i_z_NN}:
\begin{equation}
\rho_k = \sum_{i=1}^m  p_i^{\textrm{NN}} \rho_{i,k}.
\label{final_rhok}
\end{equation}
The proposed \ac{SPP} calculation is based on a \emph{hybrid} method, utilizing both the generative \ac{MoG} model and a discriminative approach to infer the posterior probability. For the latter we harness the known capabilities of the \ac{NN}.

\subsection{Training the \ac{MoG} model and the \ac{NN} classifier } \label{subsec:trainNN}
 We used the  phoneme-labeled clean speech TIMIT database~\cite{11,18} to train the \ac{NN} phoneme classifier and the \ac{MoG} phoneme-based generative model. We next   describe the training procedure.
We used the 462 speaker from the training set of the database excluding all SA sentences, since they consist of identical sentences to all speakers in the database, and hence can bias the results.

In training the phoneme-based \ac{MoG} we set the number of Gaussians to $m=39$ (see~\cite{41}), where each Gaussian corresponds to one phoneme. All frames labeled by the $i$-th phoneme were grouped, and for each frequency bin the mean and variance were computed using~\eqref{MOG} and~\eqref{ci}, respectively.
 First, the log-spectrum of the segments of clean speech utterances is calculated. Since the database is labeled each segment is associated with a phoneme $i$. We can then calculate the following first- and second-moment with phone label $i$:
\begin{equation}
	\begin{aligned}
		\mu_{i,k}&=\frac{1}{N_i}\sum_{n=1}^{N_i} x_{i,k}(n)\\
		\sigma_{i,k}^2&=\frac{1}{N_i-1}\sum_{n=1}^{N_i}\left(x_{i,k}(n)-\mu_{i,k}\right)^2
	\end{aligned}
	\label{MOG}
\end{equation}
where $x_{i,k}(n)$ is $k$-th bin of the $n$-th log-spectra vector with phoneme label $i$. The term, $N_i$ is the total number of vectors associated with phoneme labeled $i$.
The mixture coefficients $c_i$ are set to be the relative frequency of each phoneme in the training dataset:
\begin{equation}
c_i=\frac{N_i}{\sum_{n=1}^{m}N_n}.
\label{ci}
\end{equation}
Note that since the data is already labeled, no iterative clustering procedure, such as the \ac{EM} algorithm, is required.

For training the \ac{NN} as a discriminative phoneme classifier, we used the \ac{MFCC} feature vectors $\vv$ powered by the delta and delta-delta coefficients. In total, $39$ coefficients per time frame were used.   Context frames (4 from the future and 4 from the past) were added to the current frame as proposed in~\cite{9}. Hence, each time frame was represented by 351 \ac{MFCC} features. We used a single hidden layer \ac{NN} comprising of 500 neurons. (Although adding more hidden layers slightly improves phoneme classification rate, we didn't gain any significant
 improvement in the overall enhancement procedure.)  The network is constructed of sigmoid units as the transfer function for the hidden layer:
$$h_i=\frac{1}{1+\exp(-\w_{1,i}^{\T}\vv)},  \quad  i=1,\ldots,500$$ and a softmax output layer to obtain a vector $m$ probabilities associated with the various phonemes: $$p(I=i|\vv)=\frac{\exp(\w_{2,i}^{\T}\h)}{\sum_{k=1}^{m}\exp( \w_{2,k}^{\T}\h)}, \quad i=1,\ldots,m$$  where $\w_{1}$ and $\w_{2}$ are the weights matrices, of the hidden layer and the output layer, respectively.
Given a sequence of MFCC feature vectors  $\vv_1,..,\vv_N$, where $N$ is the total number of vectors in the training set, with the corresponding phoneme labels, $I_1,\ldots,I_N\in \{1,\ldots,m\} $, the \ac{NN}  is trained to maximize the log-likelihood function:
\begin{equation}
L({\w_1,\w_2}) =  \sum_{t=1}^N \log  p(I_t | \vv_t ; {\w_1,\w_2}).
\label{learnNN}
\end{equation}
To train the network we can start with random weights (or use pre-training methods (see~\cite{34})) and then, by  applying back-propagation algorithm as part of a gradient ascent procedure, the parameter sets of the network, $\w_{1}$ and $\w_{2}$, are found. In our implementation we used MATLAB$\circledR$ R2014b pattern recognition toolbox~\cite{35} to train the \ac{NN}. The default training function, namely the \emph{scaled conjugate gradient back-propagation}~\cite{36} was used. To avoid mismatch between train and test conditions each utterance was normalized, such that the utterance samples mean and the variance are zero and one, respectively.

To verify the accuracy of the classifier, the trained \ac{NN} was applied to a clean test set (24-speaker core test set drawn from TIMIT database), obtaining $71\%$  correct phoneme classification results, which is a reasonably high score.

During the test phase of the algorithm, the \ac{NN} is applied to speech signals contaminated by additive noise.  We have therefore applied the \ac{CMVN} procedure before the classifier  to circumvent the noisy test condition~\cite{12}.

 \subsection{Noise parameter initialization and adaptation} \label{subsec:noiseadapt}
  To estimate the noise parameters it is assumed that the first part of the utterance (usually 0.25~Sec) the speech is inactive and it consists of noise-only segments. These first segments can therefore be used for initializing the parameters of the noise Gaussian distribution as follows:
 \begin{equation}
\begin{aligned}
\mu_{Y,k}&=\frac{1}{N_Y}\sum_{n=1}^{N_Y} y_k(n)\\
\sigma_{Y,k}^2&={\frac{1}{N_Y-1}\sum_{n=1}^{N_Y}\left(y_k(n)-\mu_{Y,k}\right)^2}
\end{aligned}
\label{noiseinit}
\end{equation}
where $N_Y$ is the number of vectors constructed form the noise-only samples. The term $y_k(n)$ denotes the $k$-th bin of the $n$-th noise vector.

   In~\cite{3}, the noise parameters remain fixed for the entire utterance, rendering this estimate incapable of processing non-stationary noise scenarios. To alleviate this problem, we will apply an adaptation procedure (see~\cite{32} for alternative noise \ac{PSD} adaptation techniques).
 Using the \ac{SPP} derived in~\eqref{final_rhok}, the following adaptation scheme for the noise model parameters can be stated:
 \begin{equation}
  \label{noise_adapt}
	 \begin{aligned}
	 \mu_{Y,k}^{\textrm{new}}  = &\rho_k \cdot \mu_{Y,k}^{\textrm{old}} +\\ &(1-\rho_k)\left(\alpha\cdot{z_k}+(1-\alpha)\cdot \mu_{Y,k}^{\textrm{old}}\right) \\
	 \sigma_{Y,k}^{\textrm{new}}  = &\rho_k \cdot \sigma_{Y,k}^{\textrm{old}} + \\ &  (1-\rho_k) \left(\alpha\cdot\sqrt{({z_k}-\mu_{Y,k}^{\textrm{new}})^2}+(1-\alpha)\cdot\sigma_{Y,k}^{\textrm{old}}\right)
	 \end{aligned}
 \end{equation}
 where $\mu_{Y,k}^{\textrm{new}}$ and $\sigma_{Y,k}^{\textrm{new}}$ are the updated parameters and $\mu_{Y,k}^{\textrm{old}}$ and $\sigma_{Y,k}^{\textrm{old}}$ are the parameters before adaption, and $0<\alpha<1$ is a smoothing parameter. 
Using this scheme, the noise statistics can be adapted during speech utterances, utilizing the frequency bins that are dominated by noise. This scheme is particularly useful in non-stationary noise scenarios. As a consequence, the first few segments, assumed to be dominated by noise, are only used for initializing the noise statistics and their influence is fading out as more data is collected.

The proposed algorithm is summarized in Algorithm~\ref{alg:NN}.
We dub the proposed algorithm \ac{NN-MM} to emphasize its hybrid nature, as a combination of the generative MixMax model and the phoneme-classification \ac{NN}.

\begin{algorithm*}[t]
	\DontPrintSemicolon
	\underline{Train phase}:\;
    \textbf{input}: Log-spectral vectors $\z_1,\ldots,\z_N$, MFCC vectors $\vv_1,\ldots,\vv_N$, and their corresponding phoneme labels $i_1,\ldots,i_N$.\;
	{\textbf{\ac{MoG} training}}:\;
	Set the phoneme-based \ac{MoG} parameters using $(\z_1,i_1), \ldots, (\z_N,i_N)$ \eqref{MOG} and~\eqref{ci}.\;
	{\textbf{NN training}}:\;
	 Train a NN for phoneme classification using $(\vv_1,i_1), \ldots, (\vv_N,i_N)$.\;
	\underline{Test phase}:\;
	\textbf{input}: {Log-spectral vector of the noisy speech $\z$ and a corresponding MFCC vector $\vv$.}\;
	\textbf{output}: {Estimated  log-spectral vector of the clean speech $\hat{\x}$.}\;
	{Compute the phoneme classification probabilities~\eqref{p_i_z_NN}:
		$$ p_i^{\textrm{NN}}=p(I=i|\vv;\textrm{NN}), \quad i=1,\ldots,m$$
		\For{ k=1:L/2 }
		{
Compute \eqref{rhodef}:
\begin{equation*}
\rho_{i,k}^{\textrm{MM}}=p(Y_k<X_{k}|Z_k=z_k,I=i)=\frac{f_{i,k}(z_k)G_k(z_k)}{h_{i,k}(z_k)},\hspace{1cm} i=1,..,m
\end{equation*}
Compute the speech presence probability~\eqref{final_rhok}:
			\begin{equation*}
			\rho_k^{\textrm{NN-MM}} =   \sum_{i=1}^m p_i^{\textrm{NN}}  \rho_{i,k}^{\textrm{MM}}
			\end{equation*}
				Estimate the clean speech~\eqref{soft}:
				$$\hat{x}_k =  z_k - (1-\rho_{k})\cdot\beta.$$
                Adapt the noise parameters~\eqref{noise_adapt}:
  \begin{equation*}
	 \begin{aligned}
	 \mu_{Y,k}^{\textrm{new}}  = &\rho_k \cdot \mu_{Y,k}^{\textrm{old}} + (1-\rho_k)\left(\alpha\cdot{z_k}+(1-\alpha)\cdot \mu_{Y,k}^{\textrm{old}}\right) \\
	 \sigma_{Y,k}^{\textrm{new}}  = &\rho_k \cdot \sigma_{Y,k}^{\textrm{old}} +   (1-\rho_k) \left(\alpha\cdot\sqrt{({z_k}-\mu_{Y,k}^{\textrm{new}})^2}+(1-\alpha)\cdot\sigma_{Y,k}^{\textrm{old}}\right)
	 \end{aligned}
 \label{noiseadapt}
 \end{equation*}
		}
	}
\caption{Summary of the proposed \acl{NN-MM} (NN-MM) algorithm.} \label{alg:NN}
\end{algorithm*}

\section{Experimental study}\label{sec:results}
In this section we present a comparative experimental study. We first describe the experiment setup in Sec.~\ref{subsec:expsetup}. Objective quality measure results are then presented in Sec.~\ref{subsec:objectmeasurs}. In Sec.~\ref{subsec:ASR} \ac{ASR} results are compared with different approaches. Finally, the algorithm is tested with an untrained database in Sec.~\ref{subsec:notimit}.
\subsection{Experimental setup and quality measures}\label{subsec:expsetup}
To test the proposed algorithm we have contaminated speech signal with several types of noise from NOISEX-92 database~\cite{43}, namely \emph{Speech-like}, \emph{Babble}, \emph{Car}, \emph{Room}, \emph{AWGN} and \emph{Factory}.
The noise was added to the clean signal drawn from the test set of the TIMIT database (24-speaker core test set), with  5 levels of \ac{SNR} at $-5$~dB, $0$~dB, $5$~dB, $10$~dB and $15$~dB in order to represent various real-life scenarios. The algorithm was tested similarly, with the untrained  \ac{WSJ} database \cite{22}.
We  compared the proposed \ac{NN-MM} algorithm to the  \ac{OMLSA} algorithm~\cite{17} with \ac{IMCRA} noise estimator~\cite{32}, a state-of-the-art algorithm for single channel enhancement. The default parameters of the \ac{OMLSA} were set according to~\cite{40}.

In order to evaluate the performance of the \ac{NN-MM} speech enhancement algorithm, several objective and subjective measures were used, namely the \ac{PESQ} quality measure, which has a high correlation with subjective score~\cite{13}, and a composite measure~\cite{7}, weighting the \ac{LLR}, the  \ac{PESQ}  and the \ac{WSS}~\cite{48}. The composite measure  outputs \ac{Cbak}, \ac{Csig} and \ac{Covl} results.

As an additional measure we have examined the performance improvement of an \ac{ASR} system. We used the  PocketSphinx ASR system~\cite{21}. The feature set of the system is composed of 39 MFCC features powered by delta and delta-delta features. The acoustic model consists of a hidden Markov model with 5000 states. Each state is represented by a \ac{MoG} with 16 mixture components. Finally, the 20,000-word vocabulary language model was trained using \ac{WSJ} corpus~\cite{22}.
Finally, we have carried out informal listening tests\footnote{Audio samples can be found in \url{www.eng.biu.ac.il/gannot/speech-enhancement}.}.

%
%
%
%
%
%
%

\subsection{Objective results - TIMIT test set}\label{subsec:objectmeasurs}
We first evaluate the objective results of the proposed \ac{NN-MM} algorithm and compare it with the results obtained by the \ac{OMLSA} algorithm. To further examine the upper bound of the proposed method we also replaced the \ac{NN} classifier with an ideal classifier that  always provides the correct phoneme, denoted \emph{ideal-\ac{NN-MM}}.
The test set was the core set of the TIMIT database.

\begin{figure*}[tbhp]
	\centering
	\begin{subfigure}[b]{0.5\textwidth}
		\includegraphics[width=\textwidth]{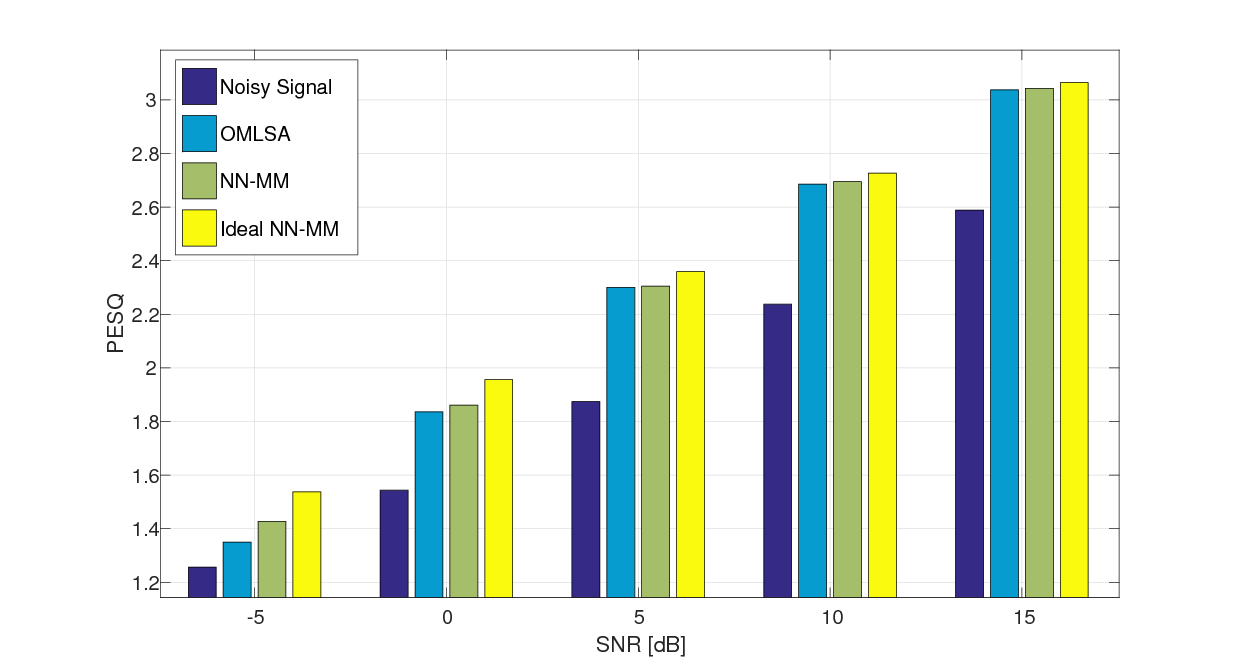}
		\caption{ {Speech} noise.}
		\label{fig:known phonemes Speech}
	\end{subfigure}%
	\begin{subfigure}[b]{0.5\textwidth}
		\includegraphics[width=\textwidth]{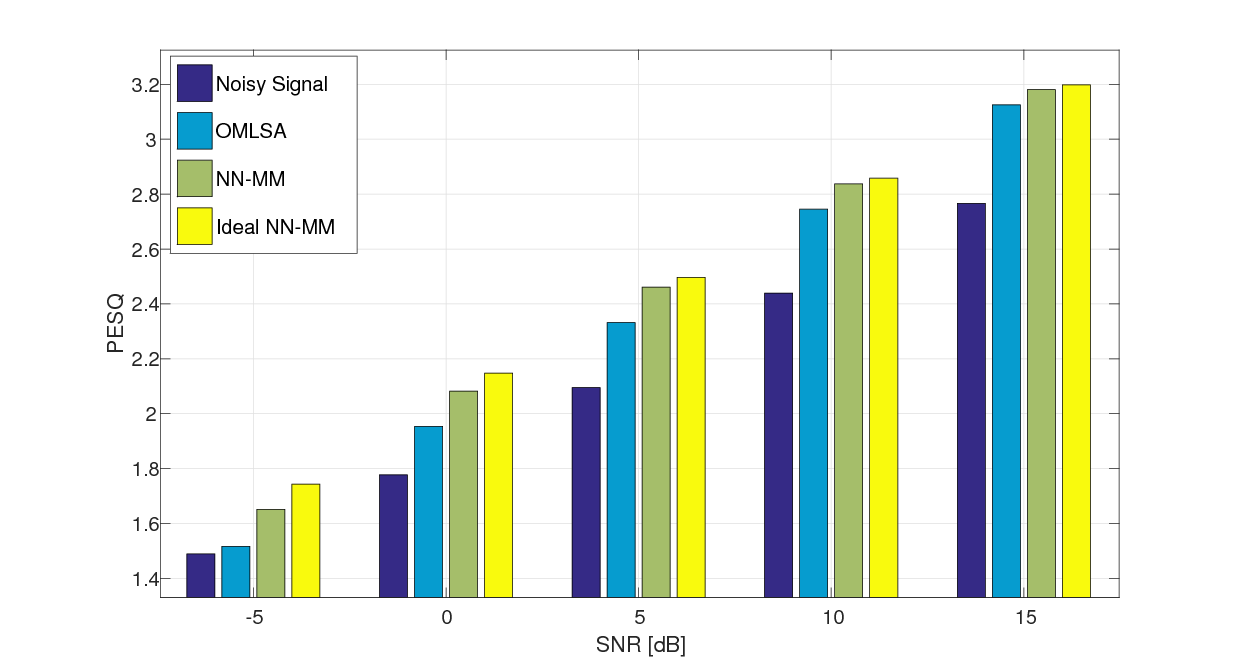}
		\caption{ {Room} noise.}
		\label{fig:known phonemes Room}
	\end{subfigure}\\		
	\begin{subfigure}[b]{0.5\textwidth}
		\includegraphics[width=\textwidth]{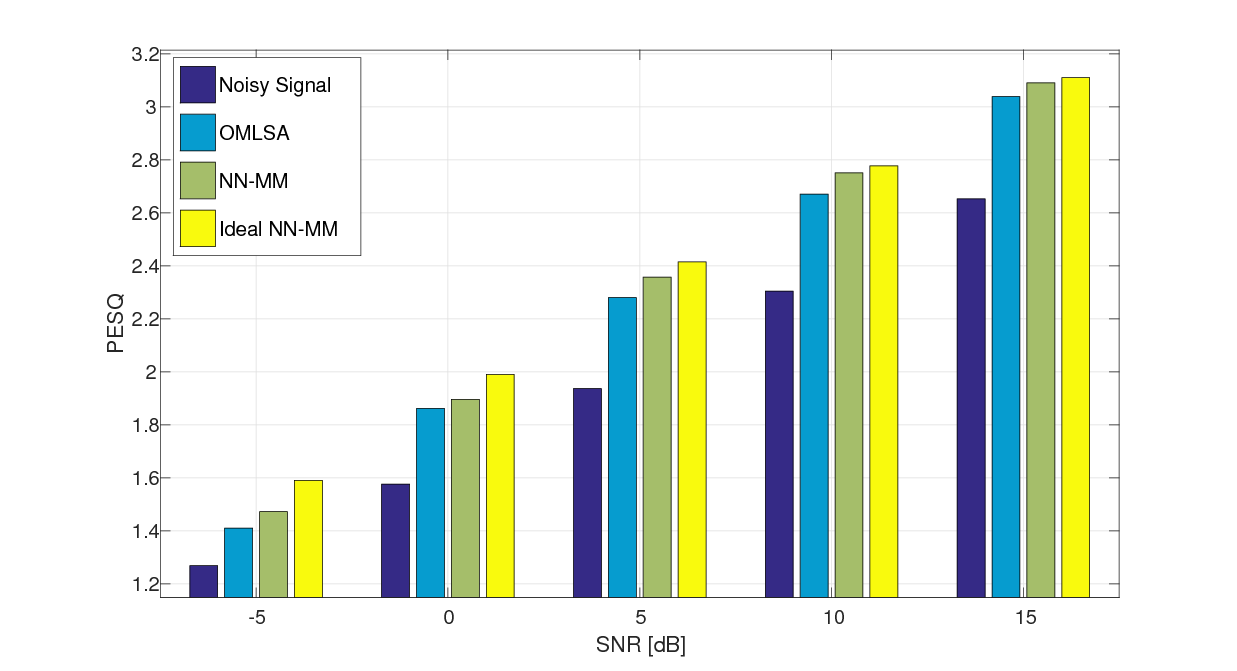}
		\caption{{Factory} noise.}
		\label{fig:known phonemes factory}
	\end{subfigure}%
	\begin{subfigure}[b]{0.5\textwidth}
		\includegraphics[width=\textwidth]{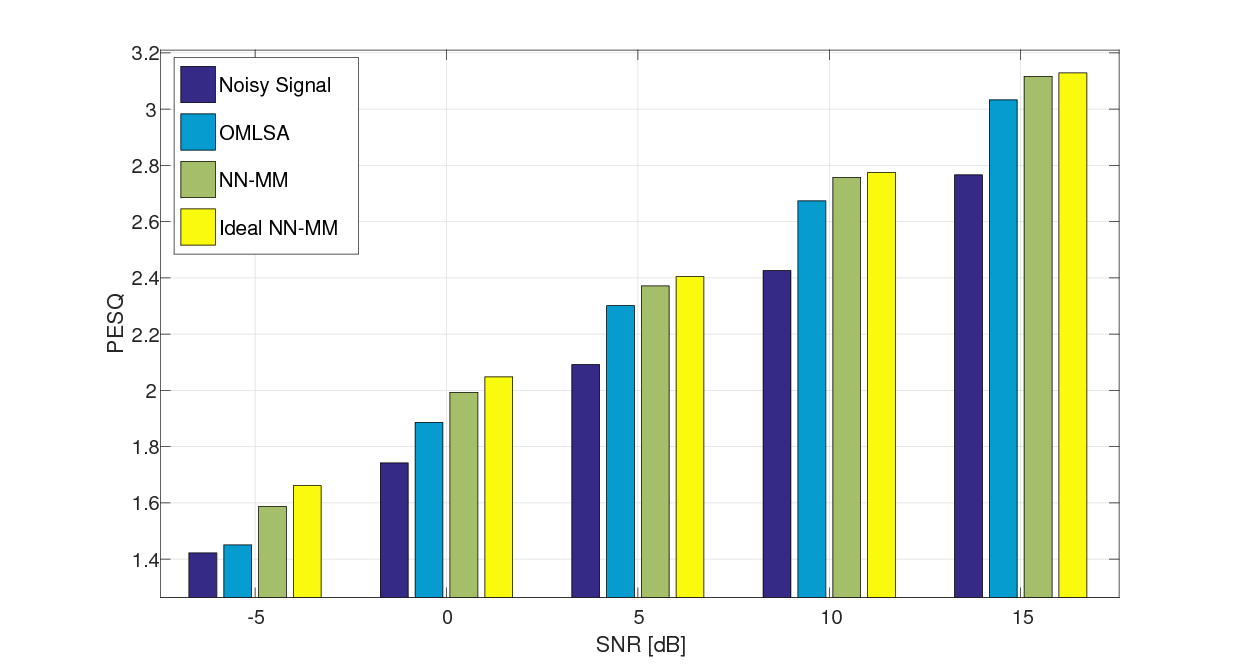}
		\caption{ {Babble} noise.}
		\label{fig:known phonemes babble}
	\end{subfigure}%
	\caption{Speech quality results (PESQ) for several noise types.}
	\label{fig:known_phonemes}
\end{figure*}

\begin{figure*}[tbhp]
	\begin{subfigure}[b]{0.5\textwidth}
		\includegraphics[width=\textwidth]{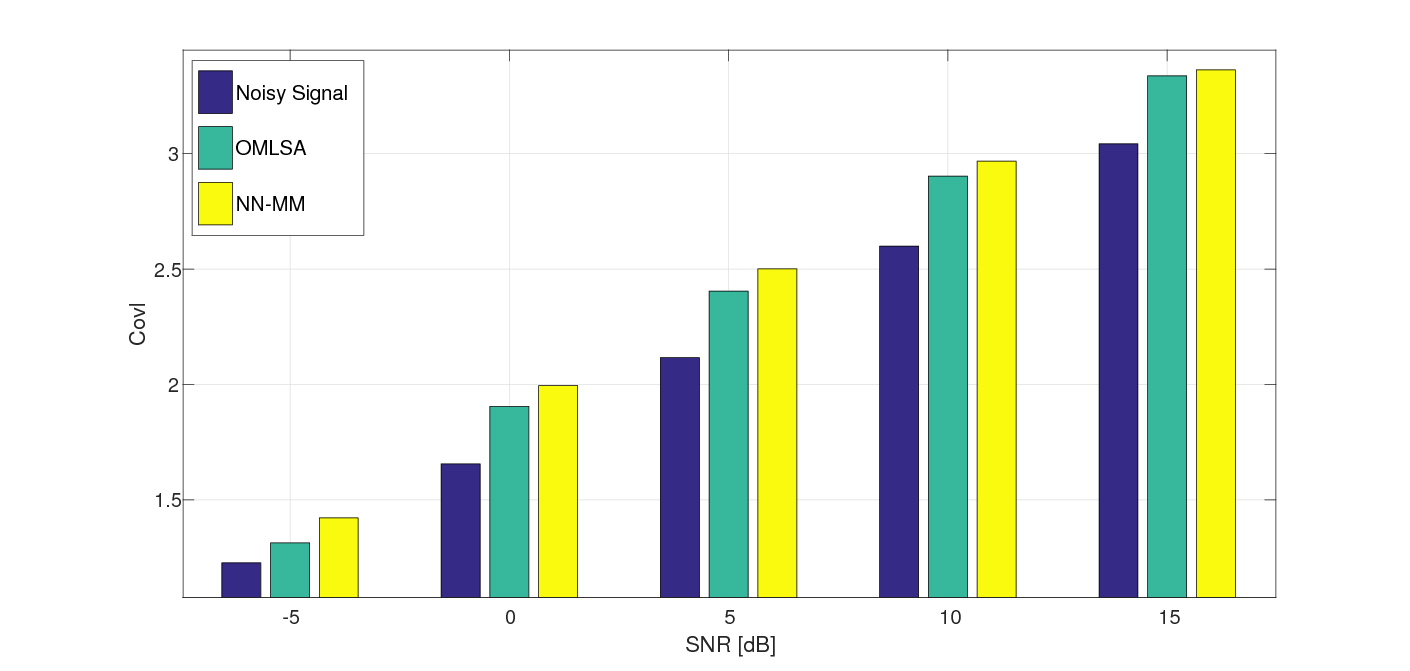}
		\caption{ {Factory} noise.}
	\end{subfigure}%
	\begin{subfigure}[b]{0.5\textwidth}
		\includegraphics[width=\textwidth]{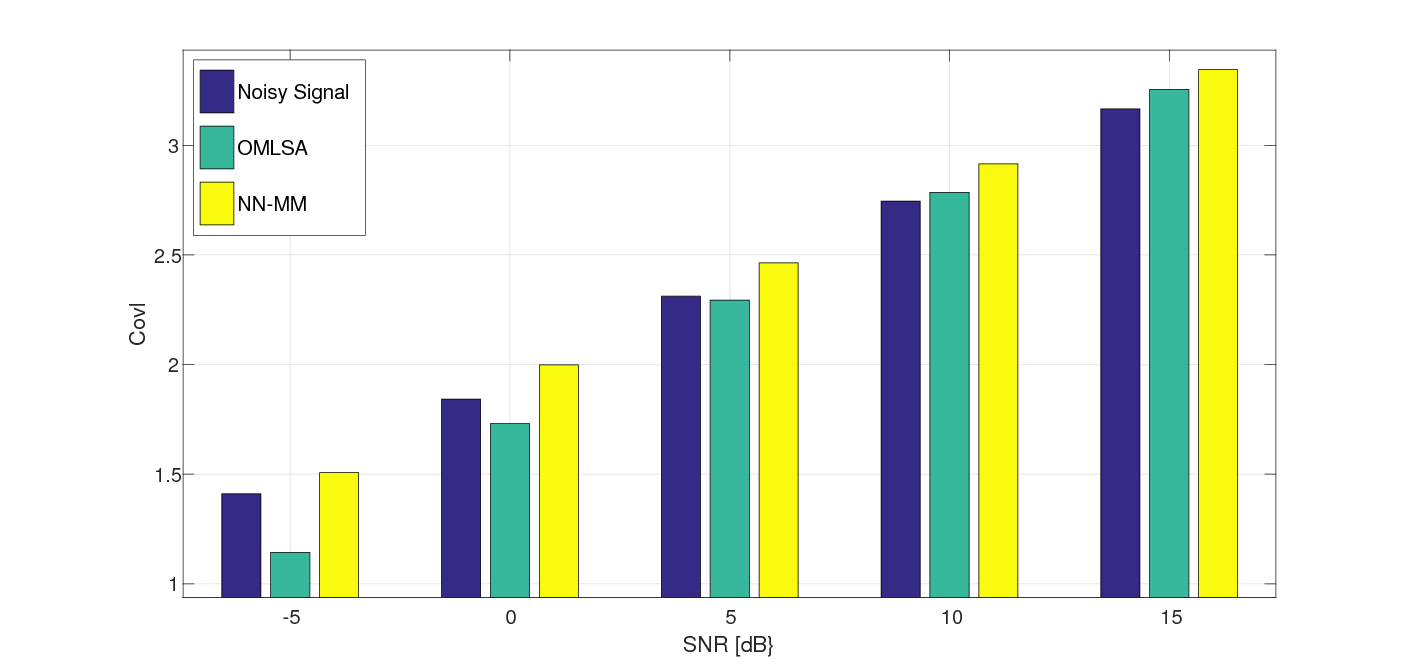}
		\caption{ {Room} noise.}
	\end{subfigure}
	\caption{Results of Covl in different noise types.}
	\label{fig:Covl}
\end{figure*}

Fig.~\ref{fig:known_phonemes} depicts the \ac{PESQ} results of all examined algorithm for the {Speech-like}, {Room}, {Factory} and {Babble} noise types as a function of the input \ac{SNR}. In Fig.~\ref{fig:Covl} we show the Covl results for factory and room noises. The results behave in a similar way,  with other noise types.

It can be clearly deduced that the proposed \ac{NN-MM} algorithm outperform the \ac{OMLSA} algorithm in the two designated objective measures. The ideal-\ac{NN-MM} outperforms the \ac{NN-MM}, but the difference is rather marginal. Still, there is a room for improvement, would a better phoneme classifier be available.

\begin{figure*}[tbhp]
		\centering
		\begin{subfigure}[normla]{0.5\textwidth}
			\includegraphics[width=\textwidth]{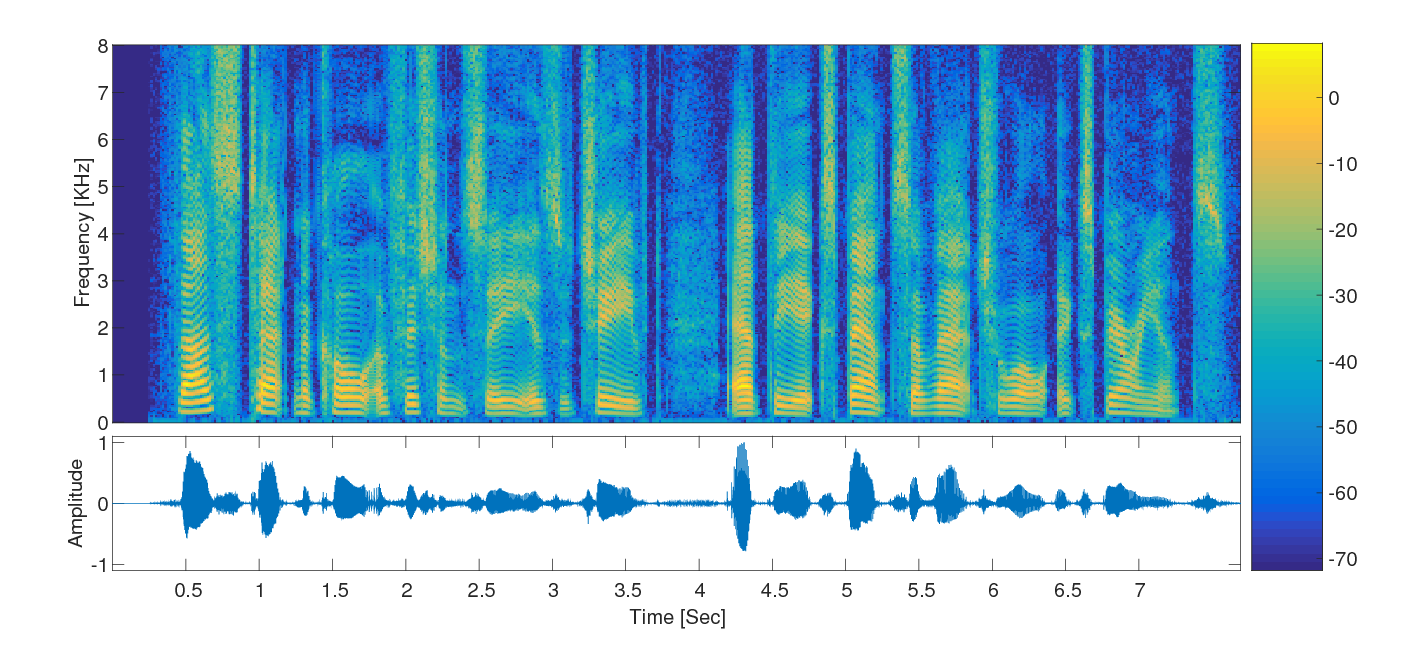}
			\caption{Clean signal.}
			\label{fig:Clean_STFT}
		\end{subfigure}%
		\begin{subfigure}[normla]{0.5\textwidth}
			\includegraphics[width=\textwidth]{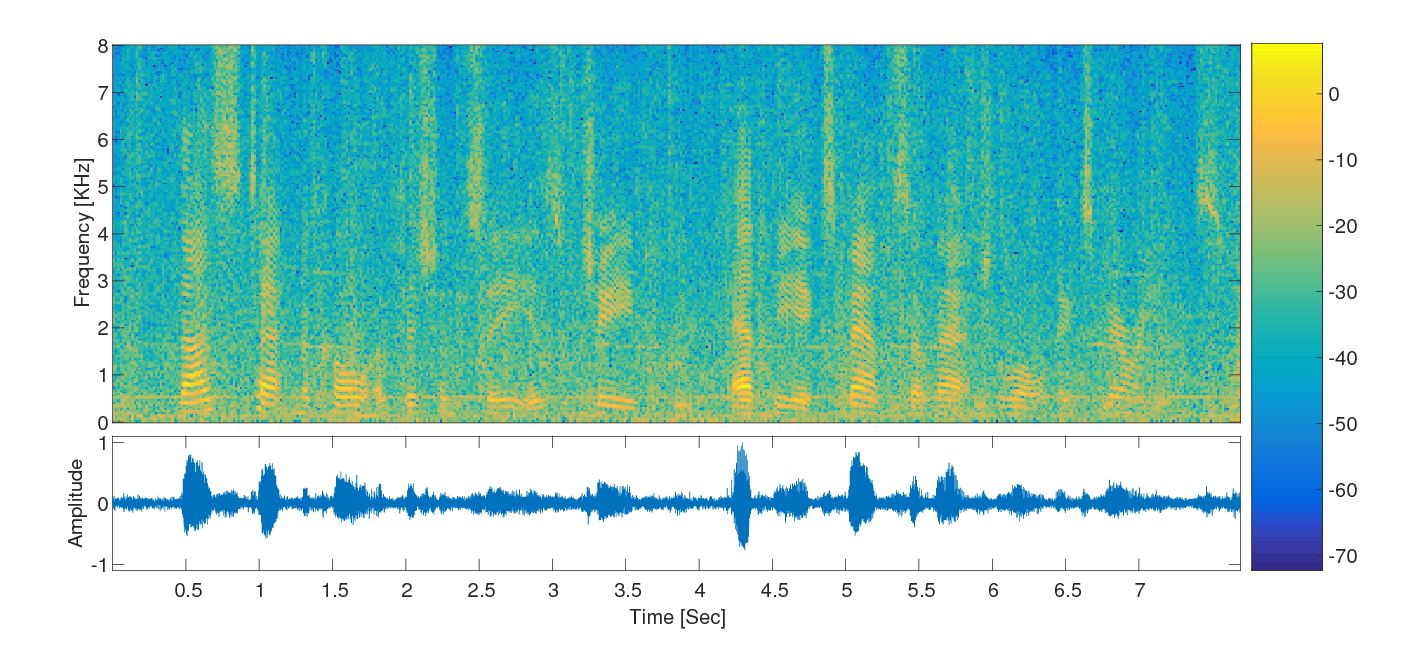}
			\caption{Noisy signal.}
			\label{fig:Noisy_STFT}
		\end{subfigure}\\
		\begin{subfigure}[normla]{0.5\textwidth}
			\includegraphics[width=\textwidth]{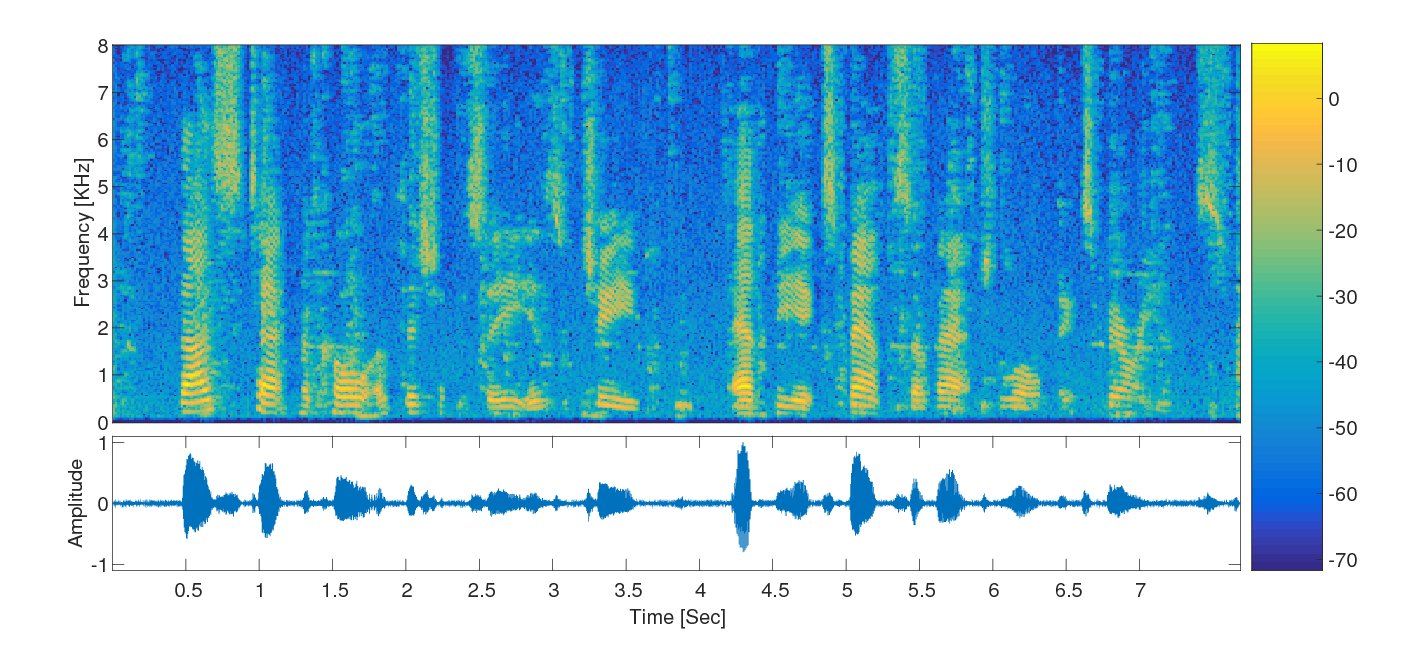}
			\caption{Signal at the output of the \ac{OMLSA} algorithm.}
			\label{fig:OMLSA_STFT}
		\end{subfigure}%
		\begin{subfigure}[normla]{0.5\textwidth}
			\includegraphics[width=\textwidth]{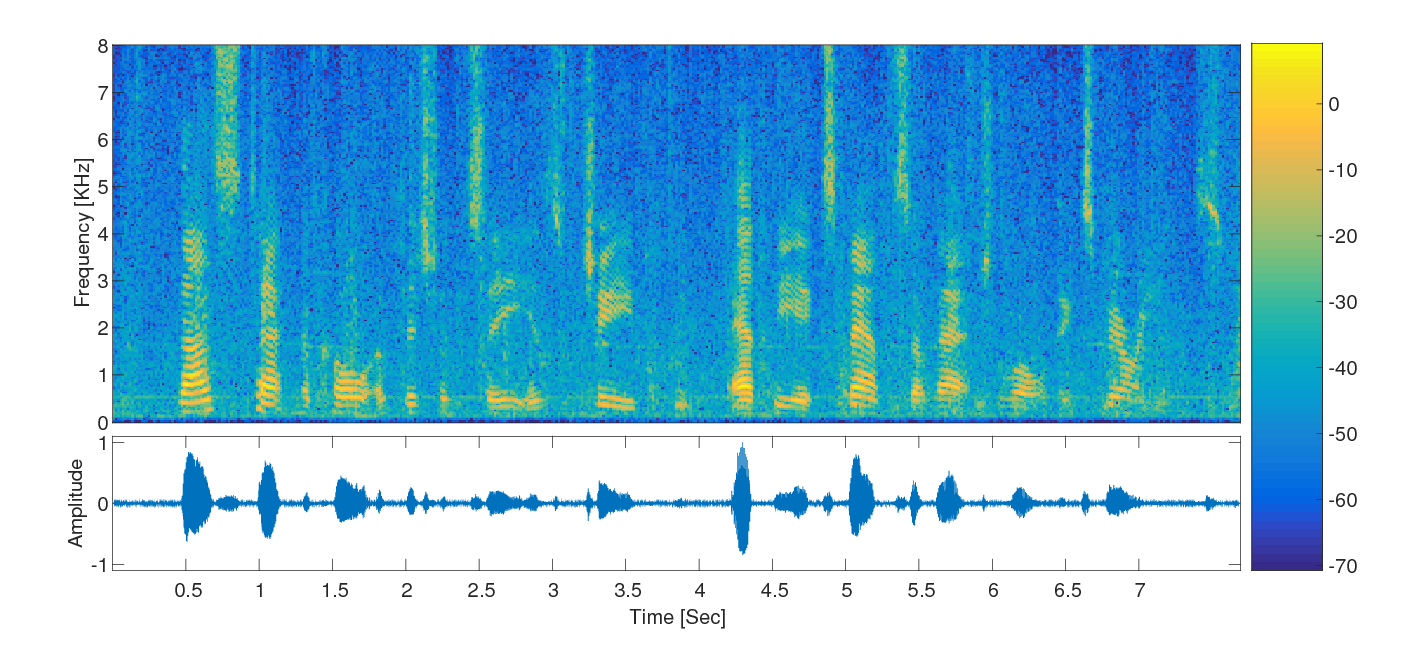}
			\caption{Signal at the output of the \ac{NN-MM} algorithm.}
			\label{fig:NNA_MM_STFT}
		\end{subfigure}
		\caption{\ac{STFT} and time-domain plots of clean, noisy (factory noise, SNR=5~dB), and signals enhanced  by the \ac{OMLSA} and the \ac{NN-MM} algorithms.}
		\label{fig:factory_Spectrum}
	\end{figure*}

	To gain further insight, we have also compared the enhancement capabilities of the proposed algorithm and the state-of-the-art \ac{OMLSA} algorithm in the challenging factory noise environment. It is clearly depicted in Fig.~\ref{fig:factory_Spectrum} (obtained in SNR=5~dB) that the proposed \ac{NN-MM} is less prone to musical noise, while maintaining comparable noise level at the output.

\subsection{Automatic speech recognition results}\label{subsec:ASR}

Speech enhancement algorithms can also serve as a preprocessing stage on front of \ac{ASR} systems.
In order to test the performance of the \ac{NN-MM} enhancement algorithm we added four types of noise in four different \ac{SNR} levels to a database comprising five female and	five male speakers, each uttering approximately 150 English sentences. The utterances were taken from different speech databases (samples from the \ac{WSJ} were not included). Overall, the test database consists of 1497 sentences, 2–4~Sec long (2–8 words each).

As before, we have the proposed \ac{NN-MM} algorithm with the original MixMax and the \ac{OMLSA} algorithms. The results are depicted in Table~\ref{table:ASR}. The \ac{NN-MM} algorithm significantly outperforms both competing algorithms for the {factory} and {babble} noise, and most the {speech-like} for most \ac{SNR} values (besides 5~dB).  In the  white noise case, the original MixMax exhibits slightly better performance. The superior results of the proposed \ac{NN-MM} algorithm can be attributed to the improved phoneme classification, which is one of the main building blocks of an \ac{ASR} system.

	\begin{table}[htbp]
		\caption{ASR results for various noise types.}
		\begin{center}
			\begin{tabular}{|l|r|r|r|r|}
				\hline
				\multicolumn{5}{|c|}     {  {Babble} noise} \\
				\hline
				Method \textbackslash	 \ac{SNR} & 5[dB] & 10[dB] & 15[dB] & 20[dB] \\ \hline
				Noisy signal & 8.8	& 43.0 &	68.8 & 79.7\\ \hline
				MixMax & 18.7 & 53.7 & 72.9	& 81.0	\\ \hline
				OMLSA &	13.7 & 45.0 & 66.0 & 76.2 \\ \hline
				NN-MM &	\textbf{28.2} & \textbf{60.3} & \textbf{76.2} & \textbf{81.9} \\
				\hline
			\end{tabular}
		\end{center}
		\begin{center}
			\begin{tabular}{|l|r|r|r|r|}
				\hline
				\multicolumn{5}{|c|}     {  {Factory} noise} \\
				\hline
				Method \textbackslash	\ac{SNR} & 5[dB] & 10[dB] & 15[dB] & 20[dB]\\ \hline
				Noisy signal & 1.1 & 32.7 & 62.2 & 76.1\\ \hline
				MixMax &	9.5 & 44.4 & 68.3 & 78.7\\ \hline
				OMLSA &	16.3 & 47.4 & 69.0 & 78.0 \\ \hline
				NN-MM &	\textbf{19.5} &\textbf{ 52.9 }& \textbf{71.9} & \textbf{80.1} \\
				\hline
			\end{tabular}
		\end{center}
		\label{table:ASR_Factory}
		\begin{center}
			\begin{tabular}{|l|r|r|r|r|}
				\hline
				\multicolumn{5}{|c|}     {  {Speech-like} noise} \\  \hline
				Method \textbackslash \ac{SNR}	& 5[dB] & 10[dB] & 15[dB] & 20[dB]\\ \hline
				Noisy signal &	7.9 &	44.4 &	68.4 &	77.5
				\\ \hline
				MixMax &	38.5 &	64.9 &	77.4 &	81.6
				\\ \hline
				OMLSA &	\textbf{41.3} &	65.4 &	75.8 &	81.3
				\\ \hline
				NN-MM &	40.4 &	\textbf{66.6} &	\textbf{78.0} &	\textbf{82.2}
				\\
				\hline
			\end{tabular}
		\end{center}
		\label{table:ASR}
		\begin{center}
			\begin{tabular}{|l|r|r|r|r|}
				\hline
				\multicolumn{5}{|c|}     {  {White} noise} \\  \hline
				Method \textbackslash	\ac{SNR} & 5[dB] & 10[dB] & 15[dB] & 20[dB]\\ \hline
				Noisy signal &	10.4 &	31.8 &	53.6 &	68.9
				\\ \hline
				MixMax &	\textbf{28.9} &\textbf{	51.7} &	\textbf{67.2} &	\textbf{77.1}
				\\ \hline
				OMLSA &	25.8 &	46.1 &	65.1 &	74.5
				\\ \hline
				NN-MM &	26.1 &	45.8 &	65.5 &	75.8
				\\
				\hline
			\end{tabular}
		\end{center}
		\label{table:ASR_White}
		
	\end{table}

	\comment{
			\begin{figure*}[th!]
				\begin{subfigure}[b]{0.5\textwidth}
					\includegraphics[width=\textwidth]{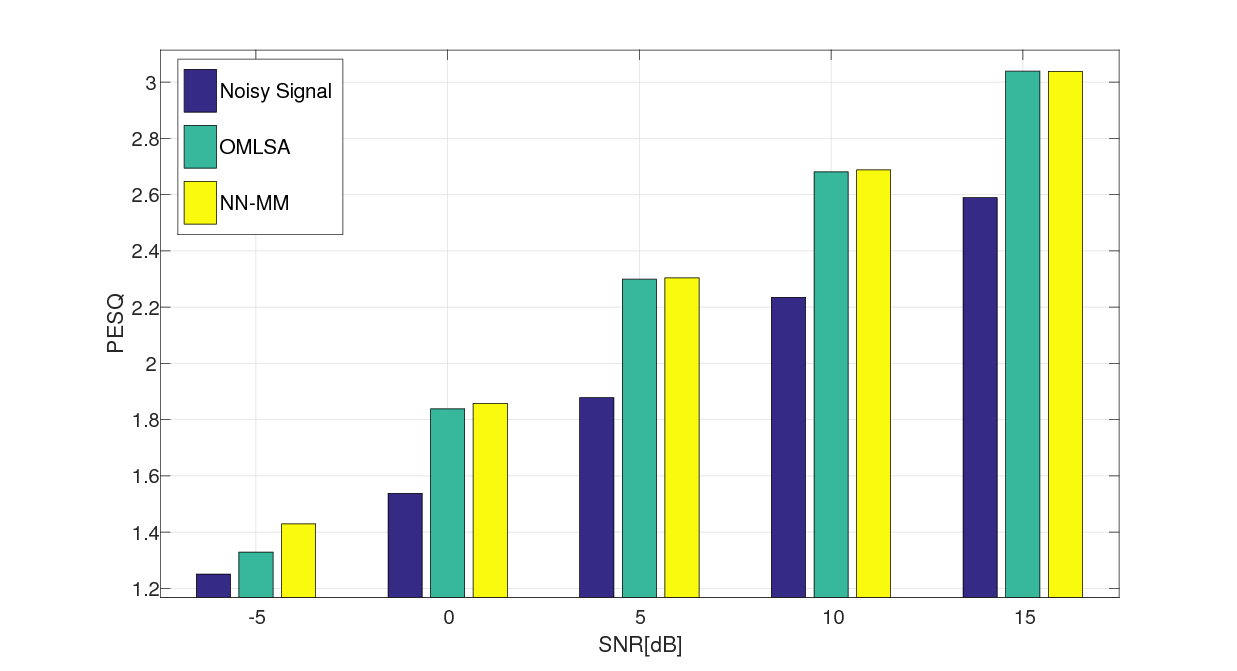}
					\caption{\emph{Speech-like} noise.}
					\label{fig:Speech_PESQ}
				\end{subfigure}%
				\begin{subfigure}[b]{0.5\textwidth}
					\includegraphics[width=\textwidth]{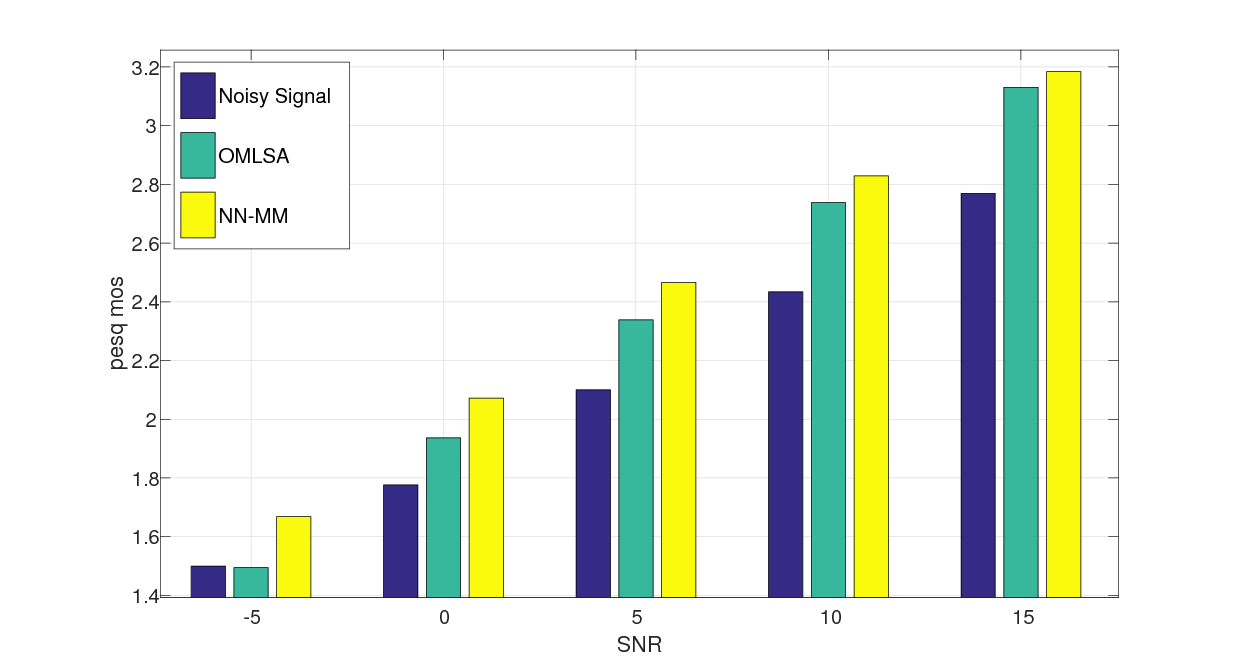}
					\caption{\emph{Room} noise.}
					\label{fig:Room_PESQ}
				\end{subfigure}\\
				\begin{subfigure}[b]{0.5\textwidth}
					\includegraphics[width=\textwidth]{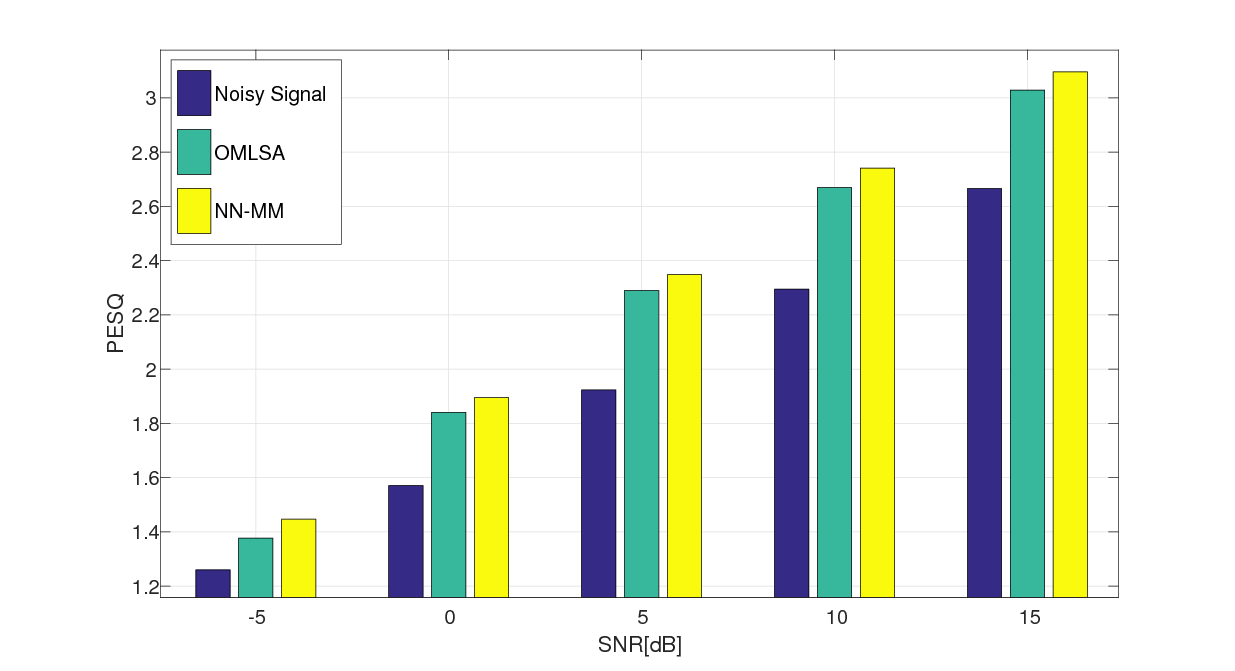}
					\caption{\emph{Factory} noise.}
					\label{fig:Factory_PESQ}
				\end{subfigure}%
				\begin{subfigure}[b]{0.5\textwidth}
					\includegraphics[width=\textwidth]{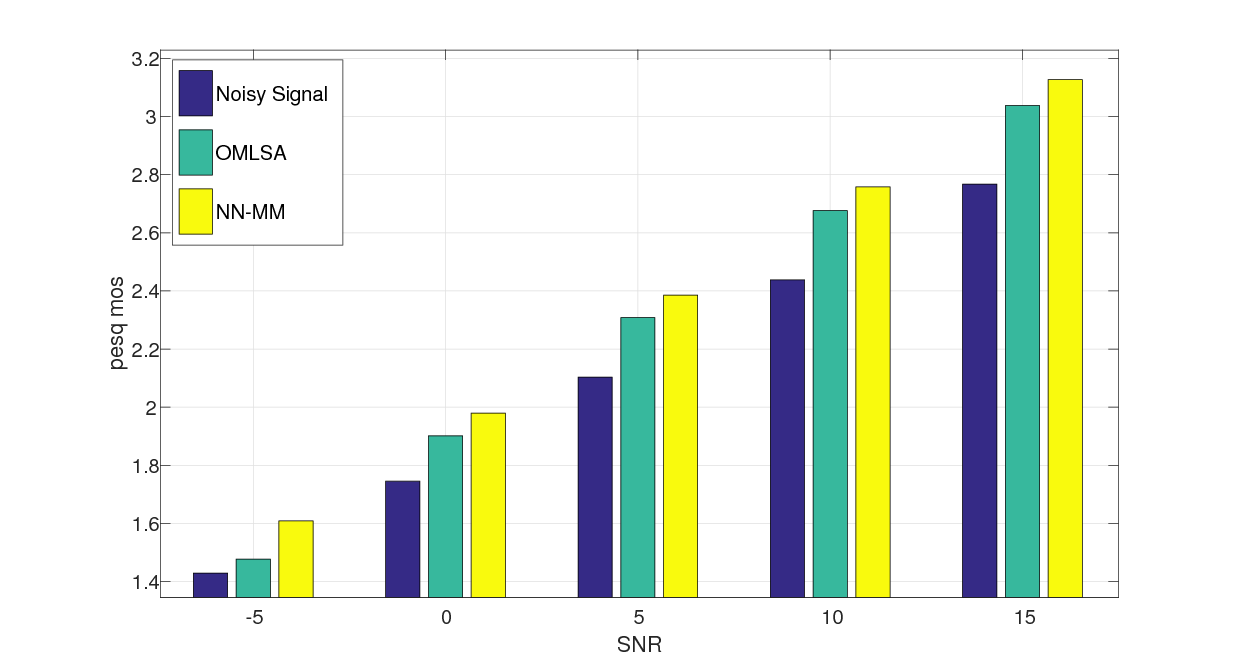}
					\caption{\emph{Babble} noise.}
					\label{fig:Babble_PESQ}
				\end{subfigure}%
				\caption{Speech quality results (PESQ) for several noise types.}\label{fig:PESQ_results}
			\end{figure*}
}

	\subsection{Performance with different database}	\label{subsec:notimit}
\begin{figure*}[tbhp]
		\begin{subfigure}[b]{0.5\textwidth}
			\includegraphics[width=\textwidth]{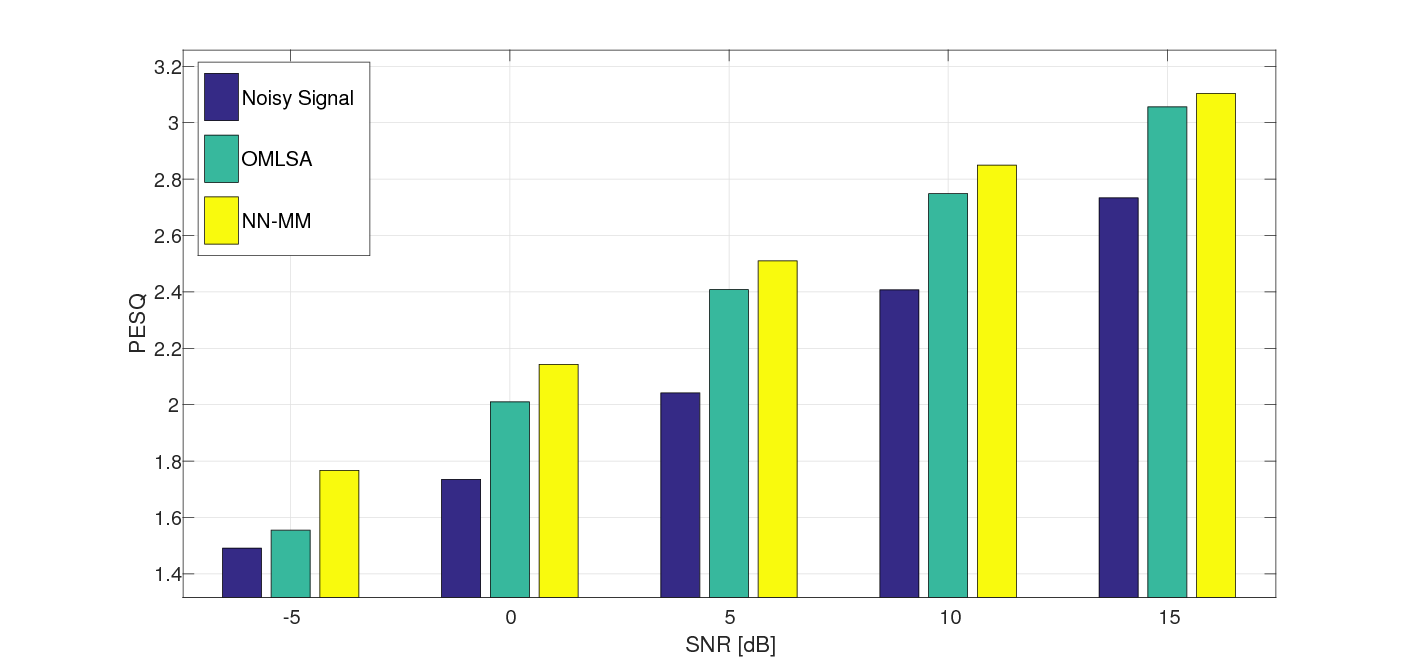}
			\caption{{Factory} noise.}
		\end{subfigure}%
		\begin{subfigure}[b]{0.5\textwidth}
			\includegraphics[width=\textwidth]{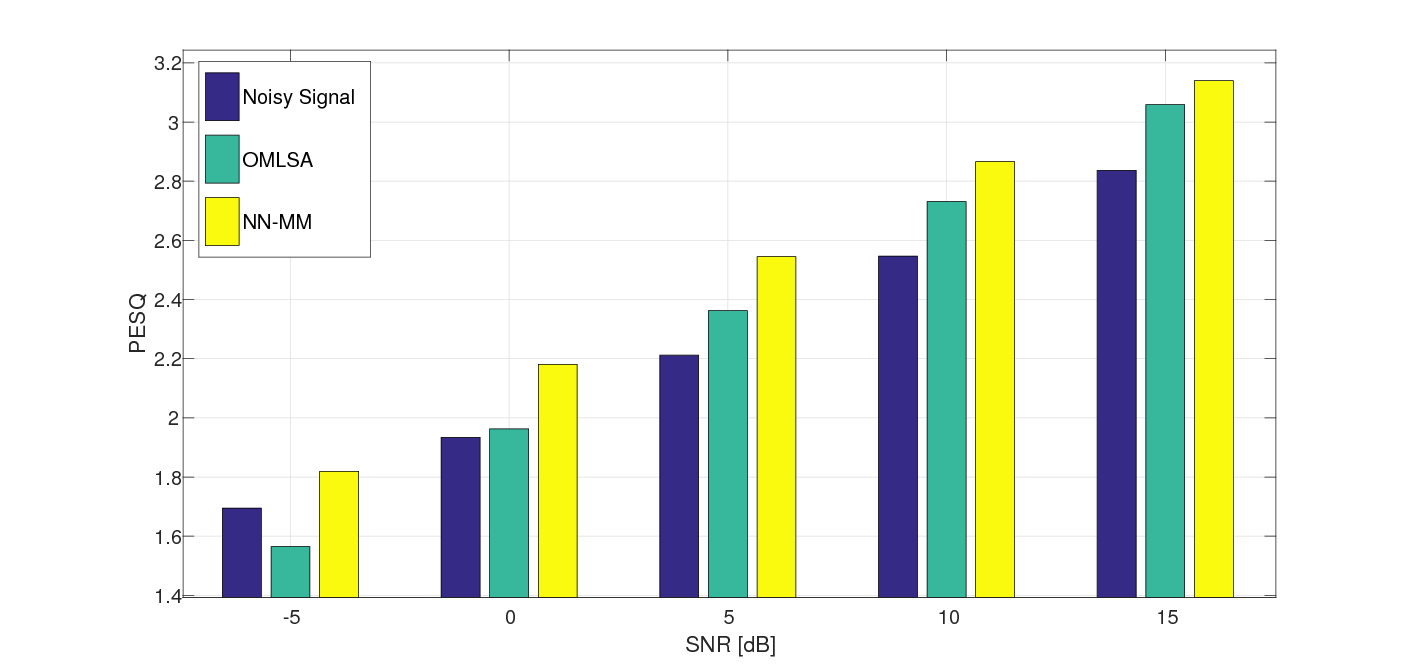}
			\caption{{Babble} noise.}
		\end{subfigure}
		\caption{PESQ results with \ac{WSJ} database for various \ac{SNR} levels.}\label{fig:No_TIMIT}
	\end{figure*}

	Finally, we would like to demonstrate the capabilities of the proposed \ac{NN-MM} algorithm when applied to speech signals from other databases. In this work we have trained the phoneme-based \ac{MoG} and the \ac{NN} using the TIMIT database. In this section we apply the algorithm to $30$ clean signals drawn from the \ac{WSJ} database~\cite{22}. The signals were contaminated by the challenging factory and babble noise with several \ac{SNR} levels. Note, that the algorithm does not train with noisy signals. Fig.~\ref{fig:No_TIMIT} depicts the \ac{PESQ} measure of the \ac{NN-MM} algorithm in comparison with the \ac{OMLSA} algorithm. It is that the performance of proposed algorithm and its advantages are maintained even for sentences from different database than the training database. Here we show the challenging factory and Babble noise types. The results in other noise types have the same construction.
	
\section{Analysis of the building blocks of the algorithm}\label{sec:analysis}
In this section, we  analyze the individual contributions of each component of the proposed algorithm to the overall performance. First, in Sec.~\ref{subsec:phonemog} the phoneme-based \ac{MoG} is analyzed.  In Sec.~\ref{subsec:spp} example of the \ac{SPP} is presented and the \ac{NN}  phoneme classifier is compared to the generative approach in Sec.~\ref{subsec:classification}. Finally, in Sec.~\ref{subsec:noiseadaptresults} the noise adaptation is tested in real-life scenario.
\subsection{The Phoneme-based \ac{MoG}} \label{subsec:phonemog}
\label{sec:MoG}
One of the major differences between the original MixMax algorithm and the proposed \ac{NN-MM} algorithm is the construction of the \ac{MoG} model. While the former uses unsupervised clustering procedure based on the \ac{EM} algorithm, the latter uses supervised clustering using the labeled phonemes. Consequently, the clusters in the proposed algorithm consists of different variants of the same phoneme, while the cluster obtained by the \ac{EM} algorithm mixtures of various phonemes. We postulate that the supervised clustering is therefore advantageous over the unsupervised clustering. We will examine this claim in the current section, using clean speech signal contaminated by {Room} noise with SNR=5~dB.

First, define the averaged \ac{PSD} of the speech utterance as the weighted sum of the Gaussian centroids, as inferred by the two clustering procedures. The weights give the respective posterior probabilities (either~\eqref{p_i_z} or~\eqref{p_i_z_NN}).
The averaged \ac{PSD} obtained by the supervised clustering and the discriminative \ac{NN} is given by:
\begin{equation}
\overline{\mu}_{k}^{\textrm{NN-MM}}=\sum_{i=1}^m p_i^{\textrm{NN}} \mu_{i,k}.
\label{avgmun_nn}
\end{equation}
Similarly, the averaged \ac{PSD} obtained by the unsupervised clustering and the generative model is given by
\begin{equation}
\overline{\mu}_{k}^{\textrm{EM}}=\sum_{i=1}^m p_{i} \mu_{i,k}^{\textrm{EM}}.
\label{avgmun_em}
\end{equation}

\begin{figure*}[tbhp]
			\centering
			\begin{subfigure}[normla]{0.5\textwidth}
				\centering
				\includegraphics[width=\textwidth]{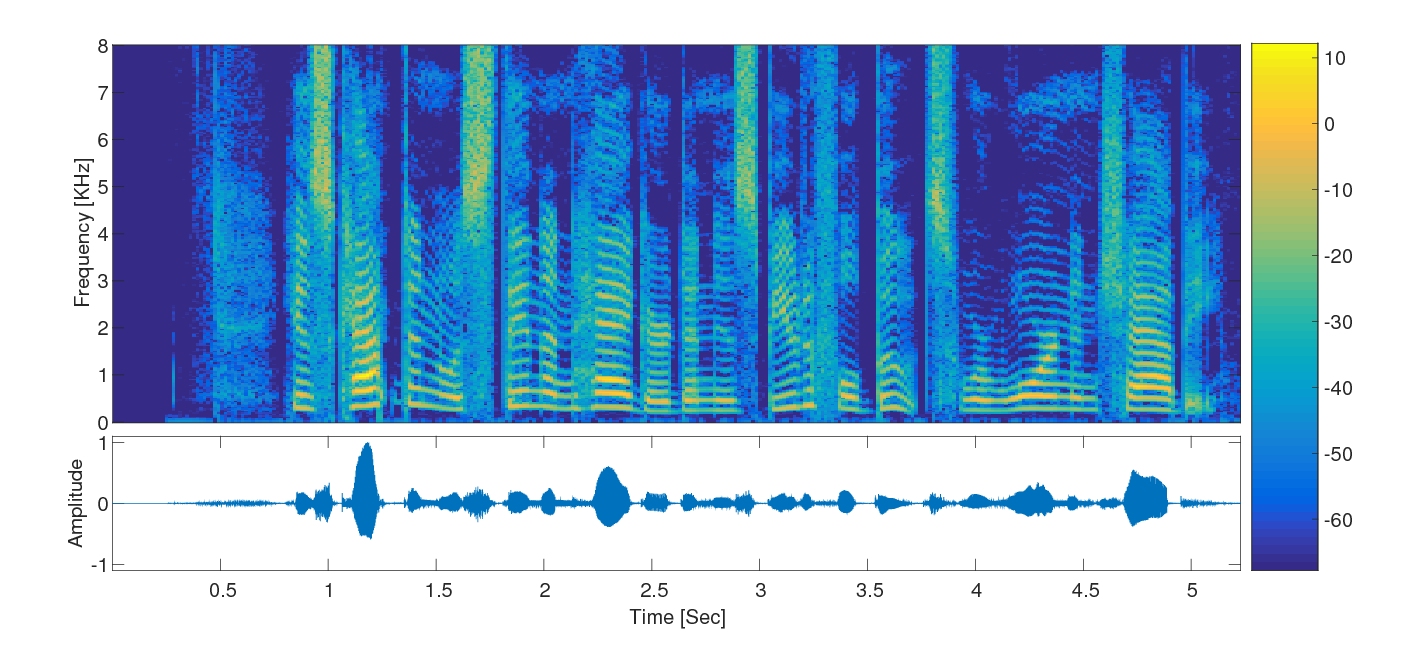}
				\caption{Clean signal}
				\label{fig:Clean_Room_5dB}
			\end{subfigure}%
			\begin{subfigure}[normla]{0.5\textwidth}
				\includegraphics[width=\textwidth]{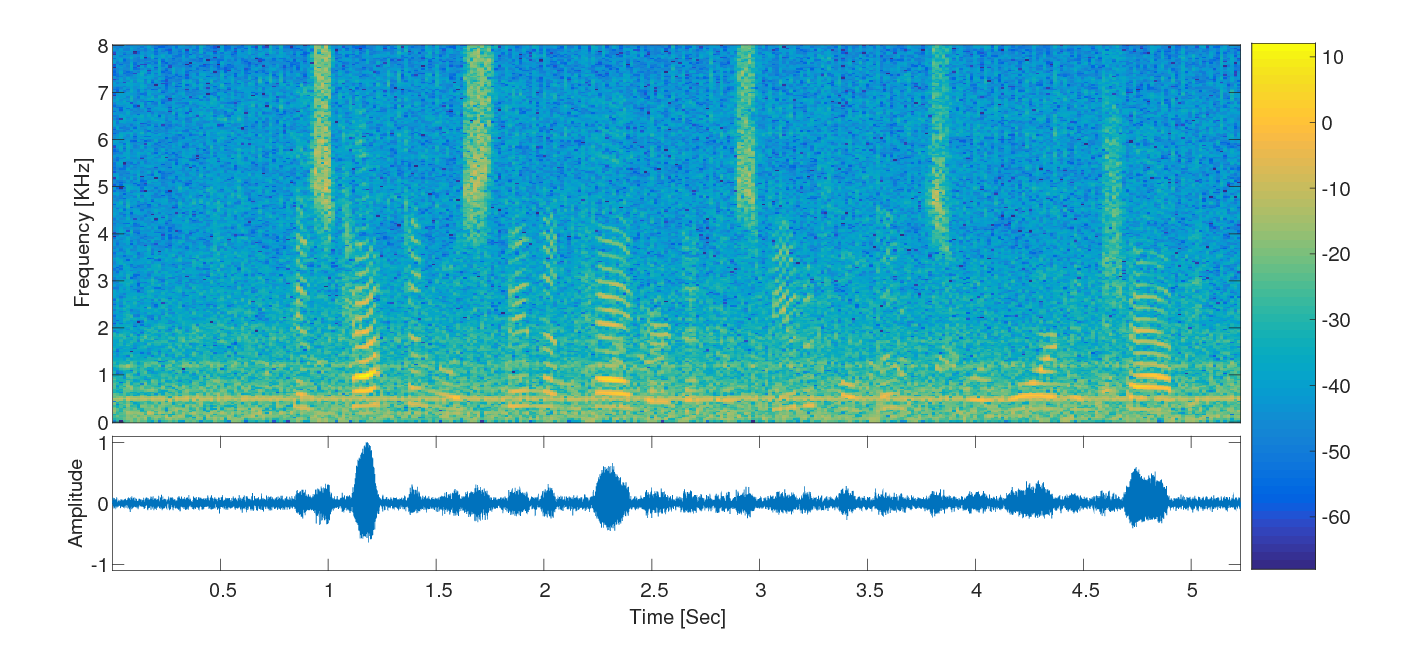}
				\caption{Noisy signal}
				\label{fig:Noisy_Room_5dB}
			\end{subfigure}%
			
			\begin{subfigure}[normla]{0.5\textwidth}
				\hspace{-0.65cm}
				\includegraphics[width=9.4cm,height=2.87cm]{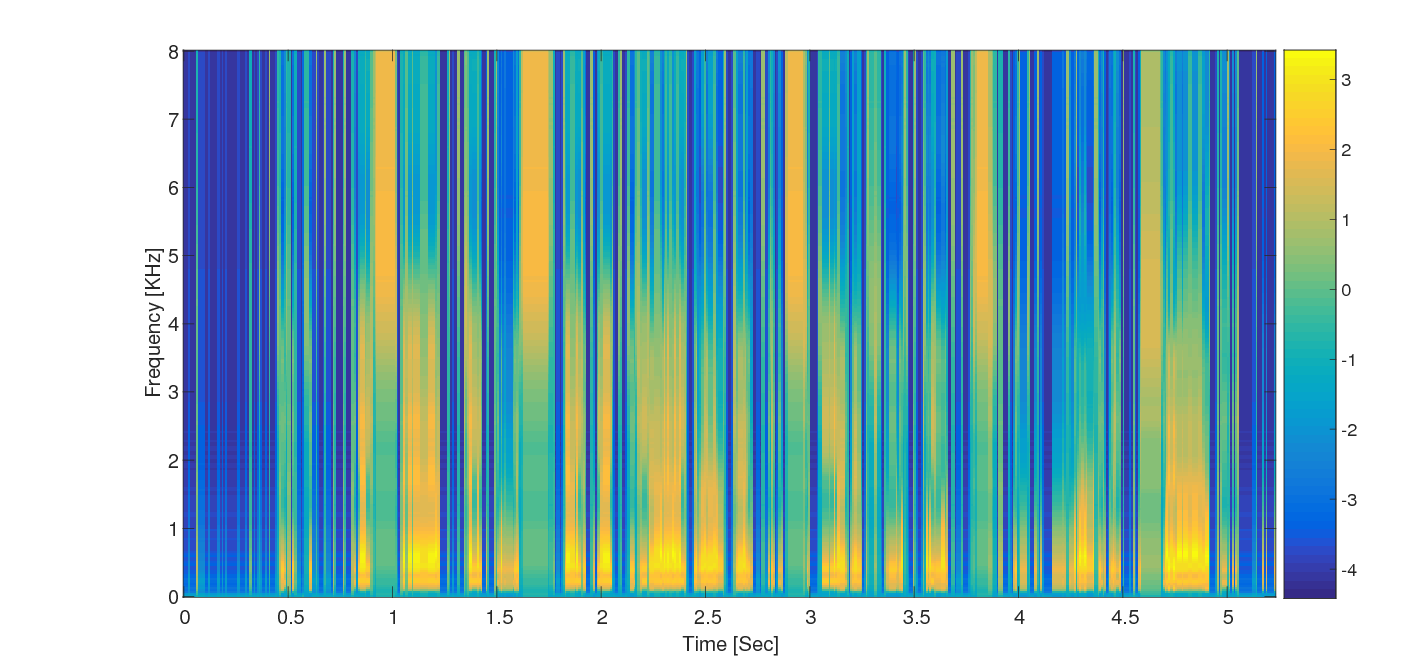}
				\caption{$\overline{\mu}^{\textrm{EM}}$ (Log-spectrum).}
				\label{fig:MUS_EM_Room_5dB}
			\end{subfigure}%
			\begin{subfigure}[normla]{0.5\textwidth}
				\hspace{-0.65cm}
				\includegraphics[width=9.4cm,height=2.87cm]{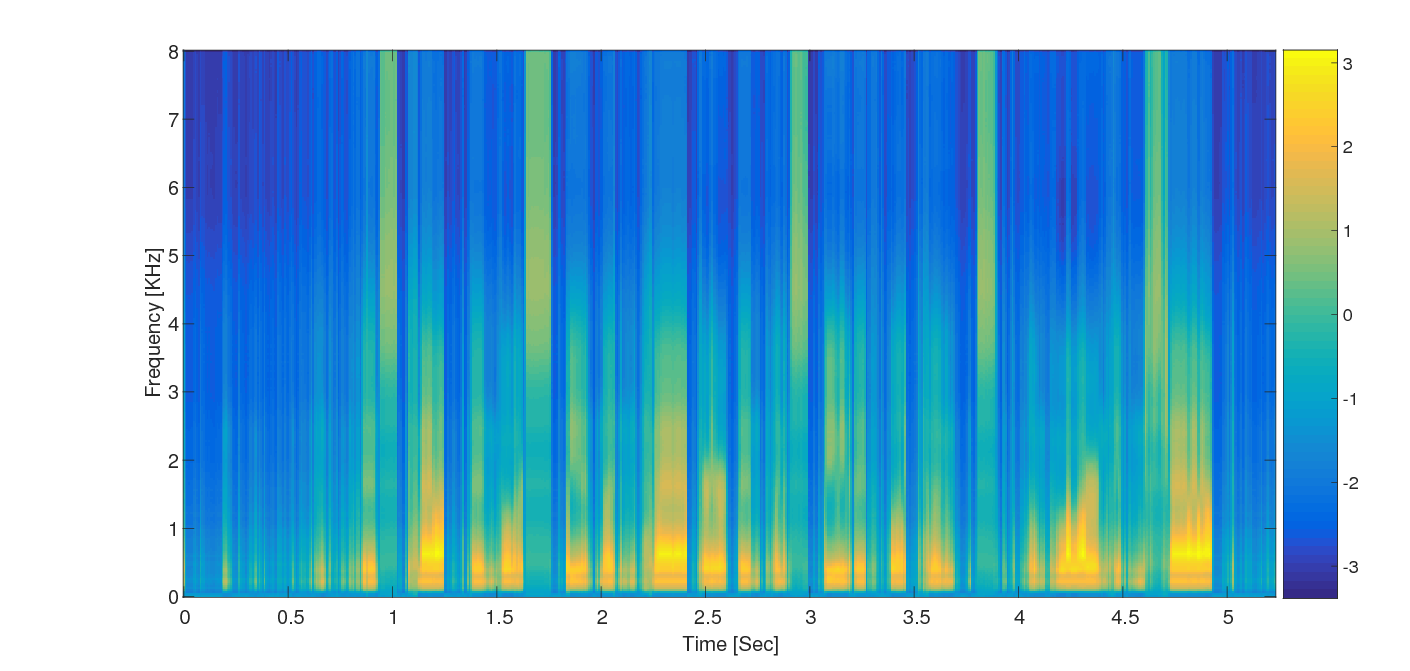}
				\caption{$\overline{\mu}^{\textrm{NN-MM}}$ (Log-spectrum).}
				\label{fig:MUS_NN_Room_5dB}
			\end{subfigure}
			
			\begin{subfigure}[normla]{0.5\textwidth}
				\hspace{-0.65cm}
				\includegraphics[width=9.4cm,height=2.87cm]{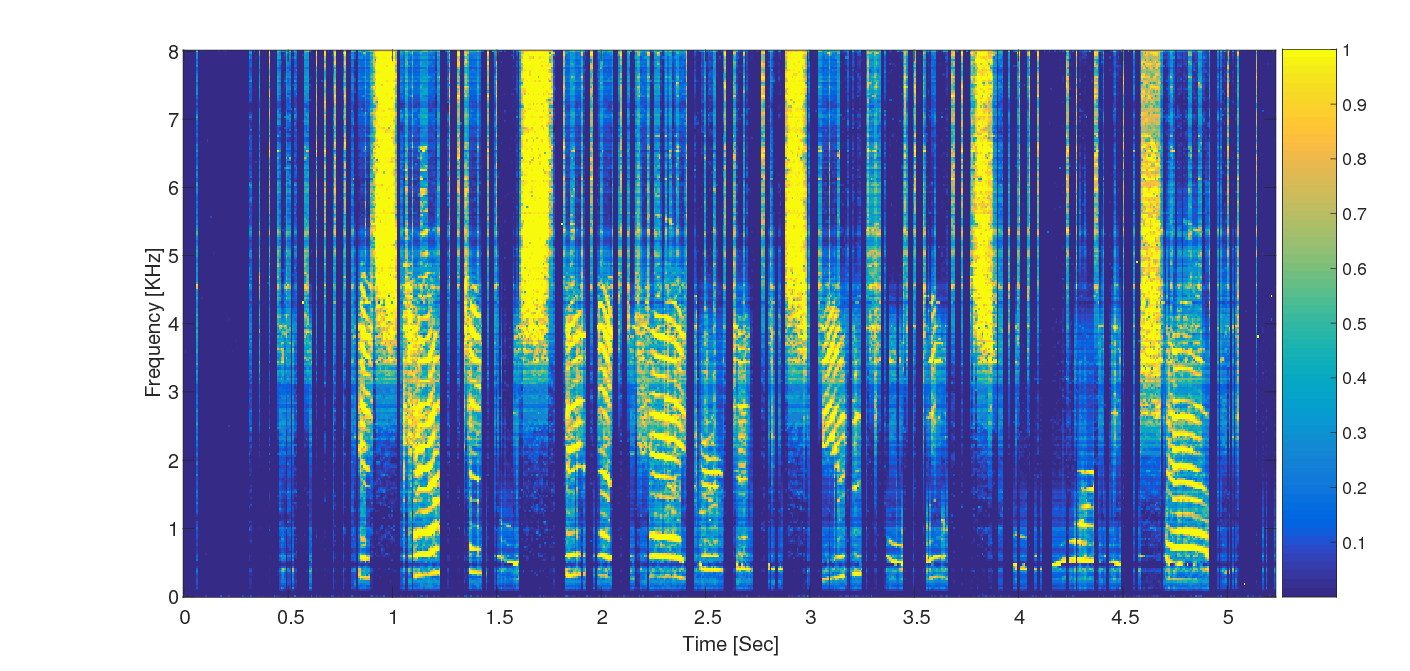}
				\caption{$\rho^{\textrm{EM}}$ parameter.}
				\label{fig:Rho_MM_Room_5dB}
			\end{subfigure}%
			\begin{subfigure}[normla]{0.5\textwidth}
				\hspace{-0.65cm}
				\includegraphics[width=9.4cm,height=2.87cm]{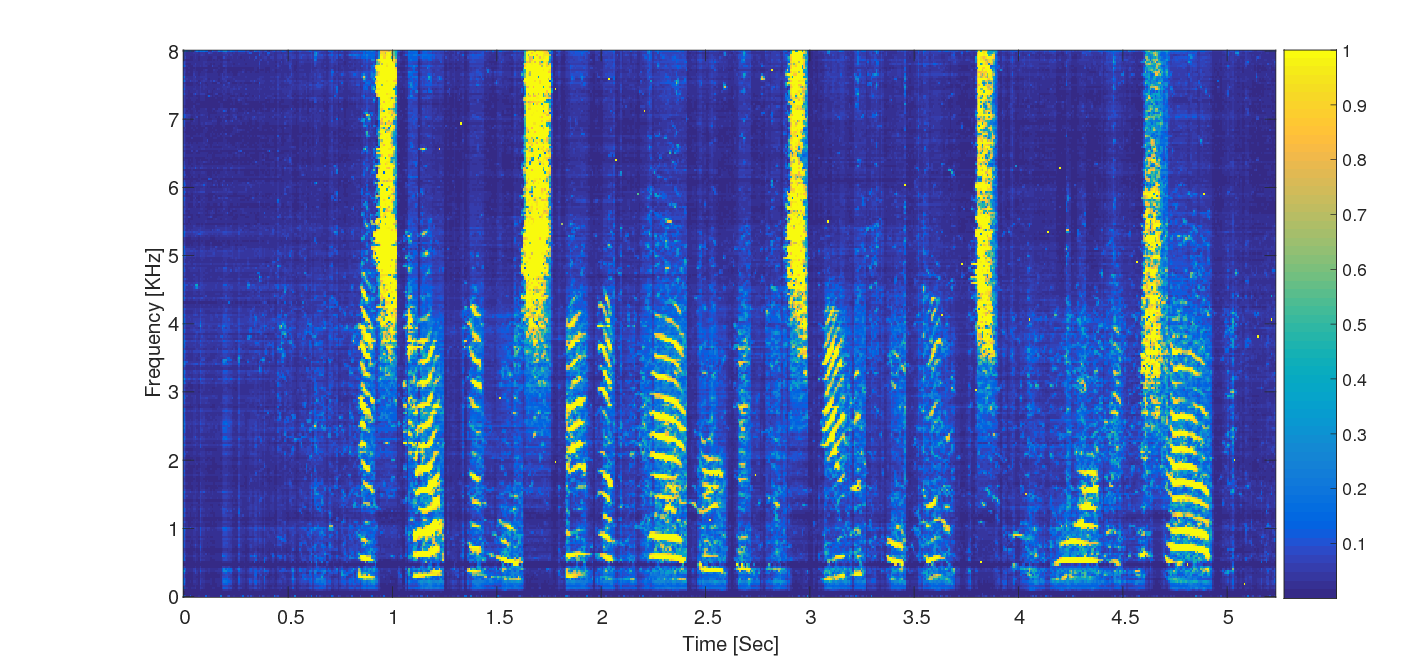}
				\caption{$\rho^{\textrm{NN-MM}}$ parameter.}
				\label{fig:Rho_NN_Room_5dB}
			\end{subfigure}
			
			\begin{subfigure}[normla]{0.5\textwidth}
				\includegraphics[width=\textwidth]{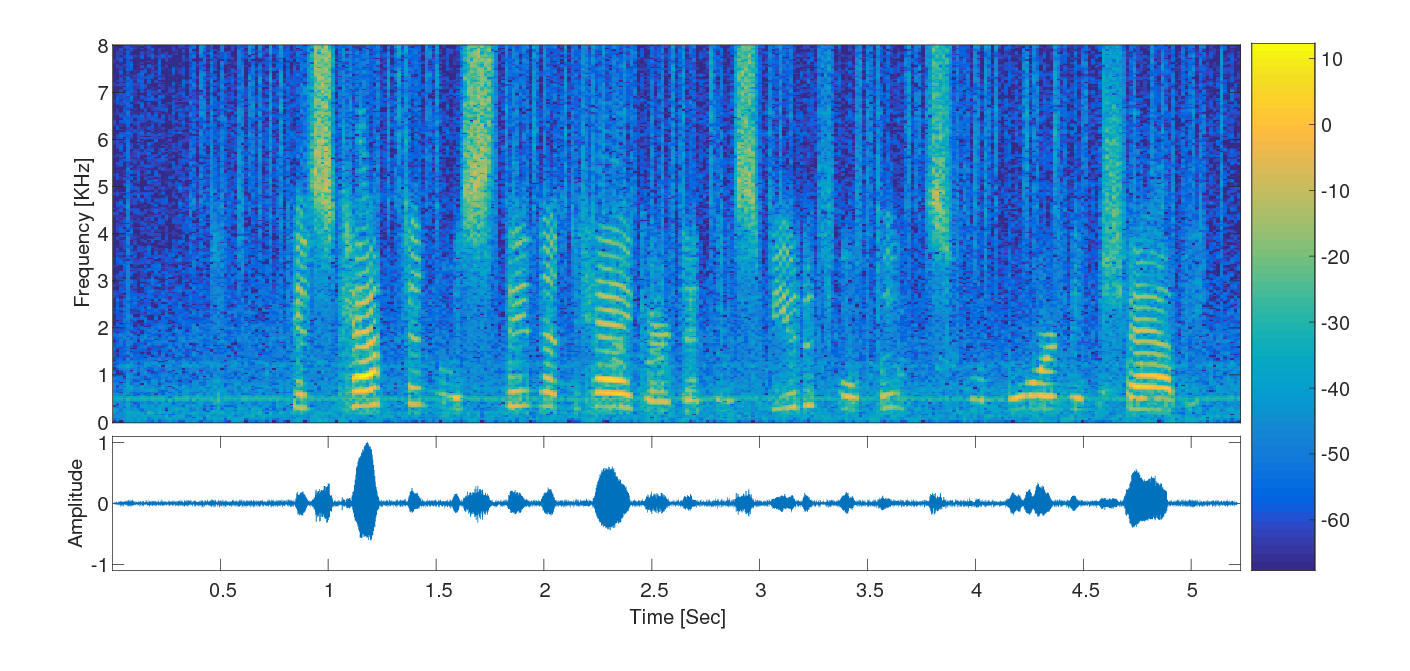}
				\caption{MixMax Enhanced.}
				\label{fig:MM_Room_5dB}
			\end{subfigure}%
			\begin{subfigure}[normla]{0.5\textwidth}
				\includegraphics[width=\textwidth]{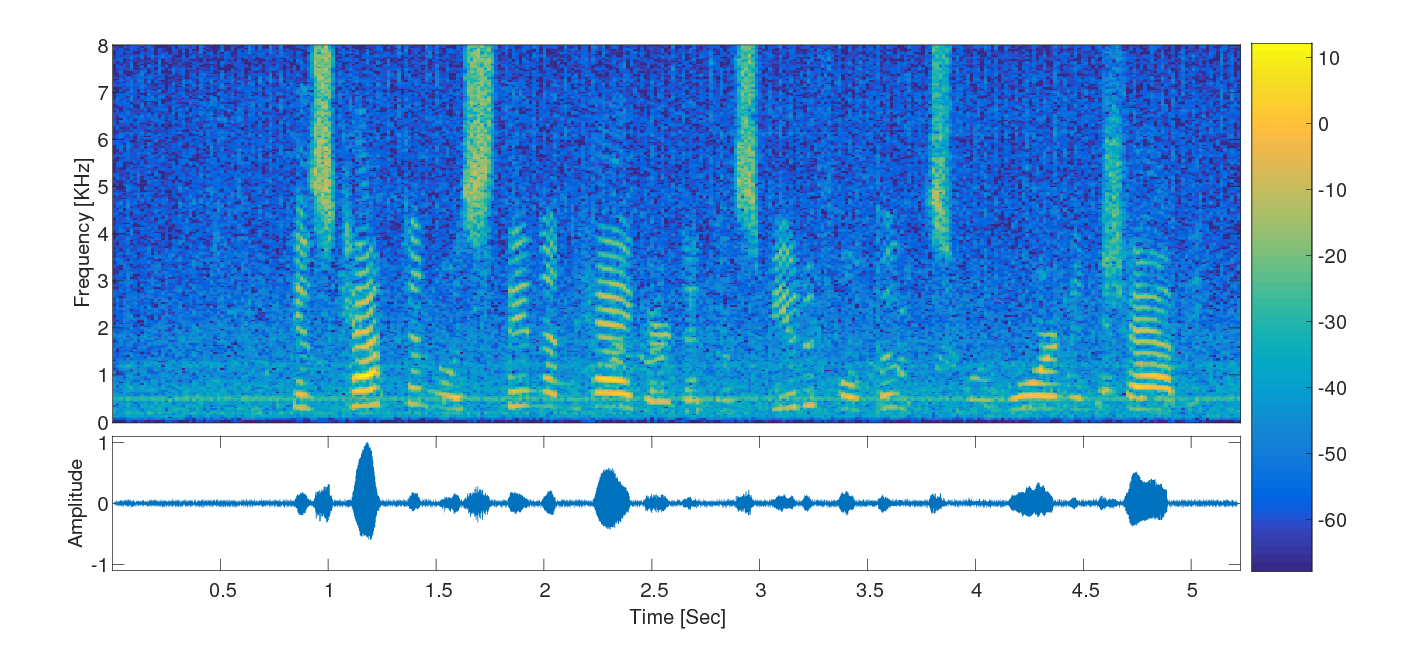}
				\caption{NN-MM Enhanced.}
				\label{fig:NN_Room_5dB}
			\end{subfigure}
			\caption{\ac{STFT} of the clean, noisy and enhanced signals together with the averaged \acp{PSD} and the \acp{SPP} using either the \ac{NN-MM} model or the original MixMax model.}
\label{fig:Spectrums&Rho}
		\end{figure*}

In Figs.~\ref{fig:Clean_Room_5dB} and~\ref{fig:Noisy_Room_5dB} we show the clean and noisy \ac{PSD}, respectively. Fig.~\ref{fig:MUS_EM_Room_5dB} and Fig.~\ref{fig:MUS_NN_Room_5dB} illustrates the estimated weighted Gaussians  $\overline{\mu}^{\textrm{NN-MM}} $ and $\overline{\mu}^{\textrm{EM}}$.  It evident that  $\overline{\mu}^{\textrm{EM}}$ is not as successful as successful as $\overline{\mu}^{\textrm{NN-MM}}$ in estimating the clean speech \ac{PSD}.

\subsection{The \acl{SPP}}\label{subsec:spp}
The \ac{SPP} is  the probability that the time-frequency bin is dominated by speech. In this section we examine the \ac{SPP} $\rho_k^{\textrm{NN-MM}}$ developed in this work as given in Algorithm~\ref{alg:NN}.
To further validate the advantages of the hybrid scheme we compare it with the \ac{SPP} used in the original MixMax, namely the posterior probabilities are inferred from the generative model and the \ac{MoG} is trained in an unsupervised manner. The latter \ac{SPP} is denoted $\rho^{\textrm{EM}}$

We continue the example in Sec.~\ref{sec:MoG}. Both \acp{SPP}, $\rho^{\textrm{EM}}$ and $\rho^{\textrm{NN-MM}}$, are depicted in Figs.~\ref{fig:Rho_MM_Room_5dB} and~\ref{fig:Rho_NN_Room_5dB}, respectively. It can be easily observed that $\rho^{\textrm{NN-MM}}$ has a better resemblance to the clean speech spectrogram shown in Fig.~\ref{fig:Clean_Room_5dB} and suffers from less artifacts. Additionally, it is smoother than the $\rho^{\textrm{EM}}$ in both time and frequency aspects. Conversely, vertical narrow spectral lines can be easily observed in $\rho^{\textrm{EM}}$. This spectral artifacts may be one of the causes for the differences in the enhancement capabilities of the two algorithms, as depicted in Figs.~\ref{fig:Rho_MM_Room_5dB} and~\ref{fig:Rho_NN_Room_5dB}.

We postulate that the designated advantages of the proposed approach stem from the better classification capabilities as exhibited by the \ac{NN}. While the original MixMax algorithm is only utilizing the current frame for inferring the posterior probabilities, the proposed algorithm takes into account the context of the phoneme by augmenting past and future frames to the current frame. This guarantees a smoother \ac{SPP} and consequently less artifacts at the output of the algorithm.

This context-aware feature vector together with the phoneme-based \ac{MoG} may also alleviate the musical noise phenomenon. This observation is also supported by the smoother spectral envelop of the \ac{MoG} centroid as can be deduced from comparing Figs.~\ref{fig:MUS_EM_Room_5dB} and~\ref{fig:MUS_NN_Room_5dB}.

\subsection{Phoneme classification task}\label{subsec:classification}
We turn now to the assessment of the proposed phoneme classifier. For that we compare the classification accuracy of the \ac{NN} with that of the generative model in~\eqref{p_i_z} using the phoneme-labeled \ac{MoG}.

\begin{figure*}[tbhp]
	\begin{subfigure}[b]{0.5\textwidth}
		\includegraphics[width=\textwidth]{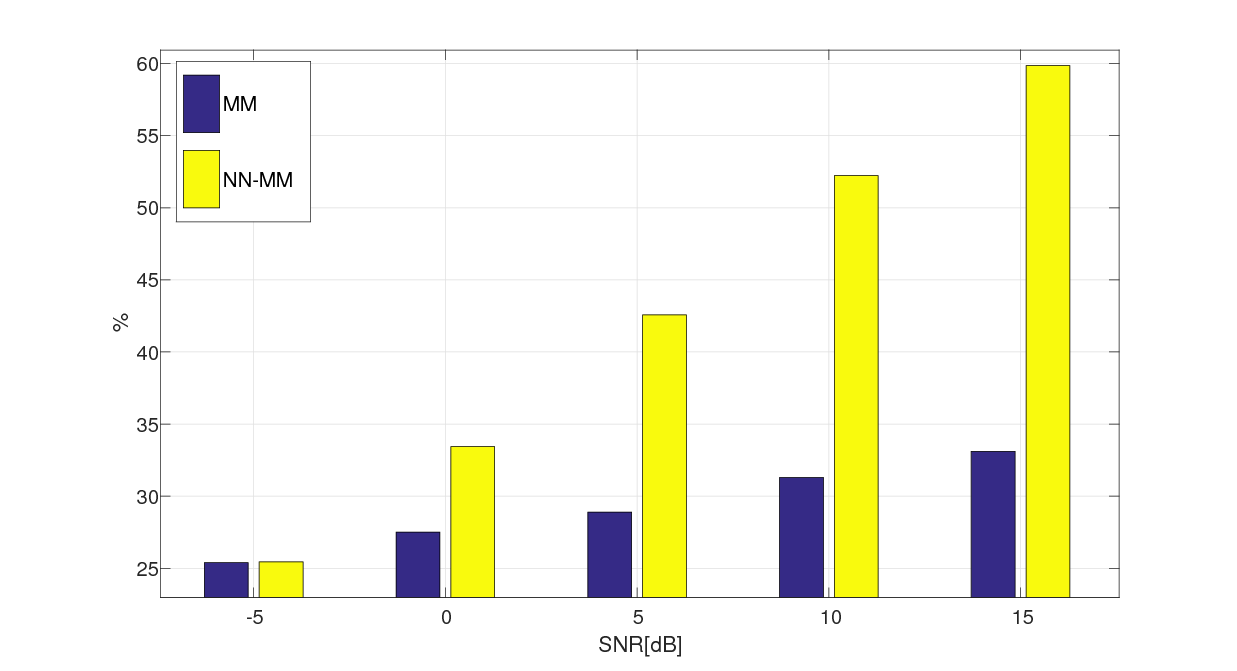}
		\caption{ {Factory} noise.}
	\end{subfigure}%
	\begin{subfigure}[b]{0.5\textwidth}
		\includegraphics[width=\textwidth]{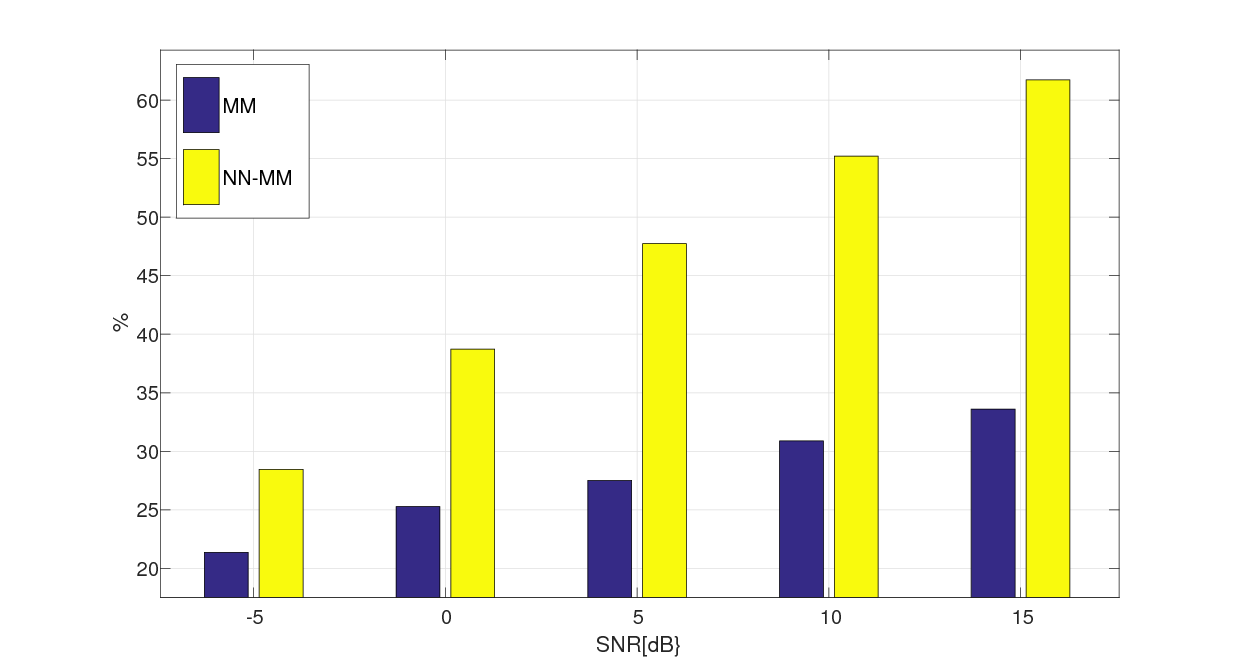}
		\caption{ {Babble} noise.}
	\end{subfigure}
	\caption{Results of phoneme classification task performed on  noisy data. }\label{fig:Classification_results}
\end{figure*}

Fig.~\ref{fig:Classification_results} depicts the percentage of correct classification results obtained on the test data. Two types of noise were added to the clean signals, namely factory and babble noise. The results clearly indicate that the \ac{NN} based classifier significantly outperforms the classifier based on the generative model, and hence better suited for the task at hand.

\begin{figure*}[tbhp]
\centering
\begin{subfigure}[normla]{0.5\textwidth}
\centering
\includegraphics[width=\textwidth]{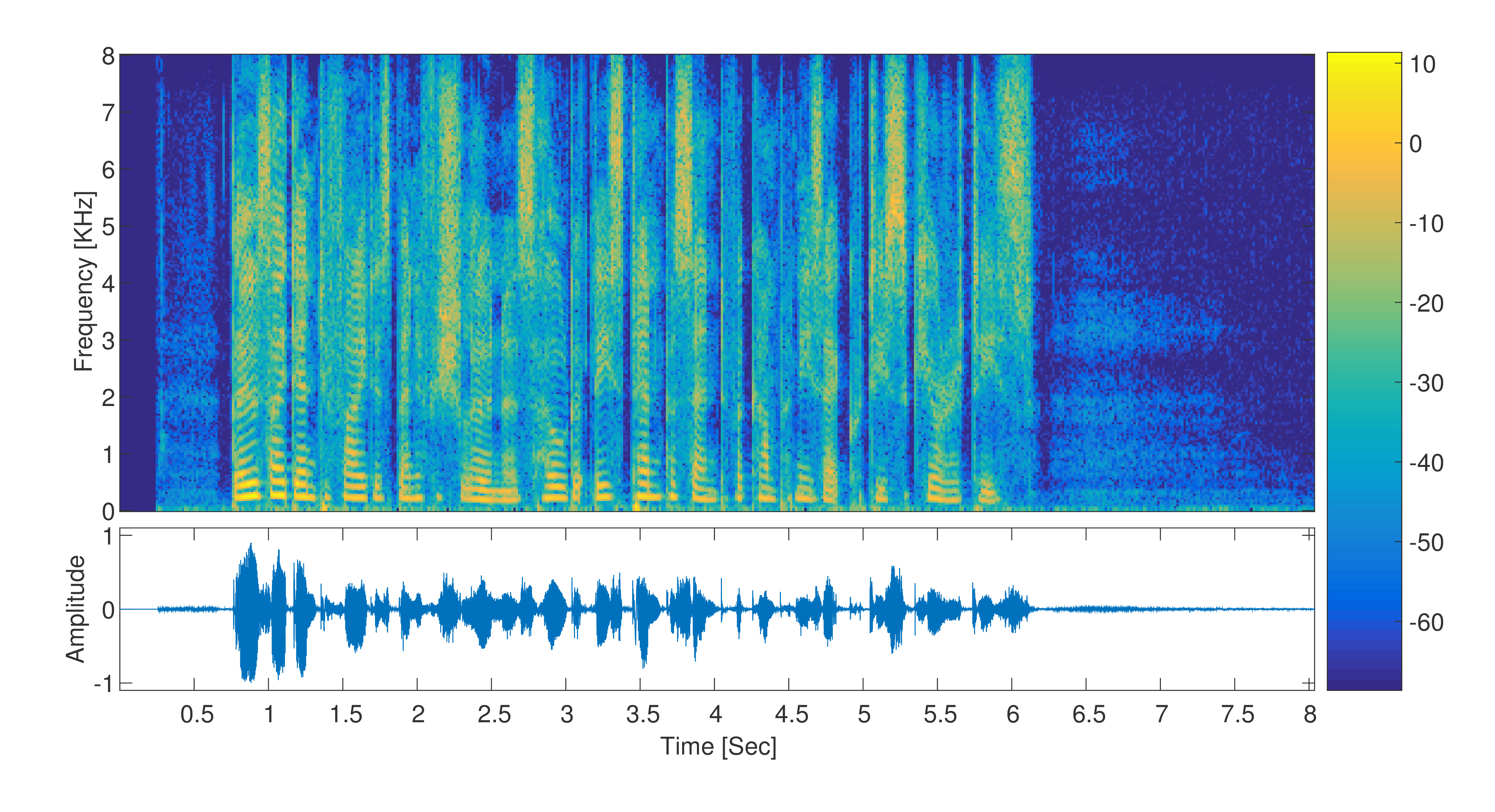}
\caption{Clean signal.}
\label{fig:Clean_siren_STFT}
\end{subfigure}%
\begin{subfigure}[normla]{0.5\textwidth}
\includegraphics[width=\textwidth]{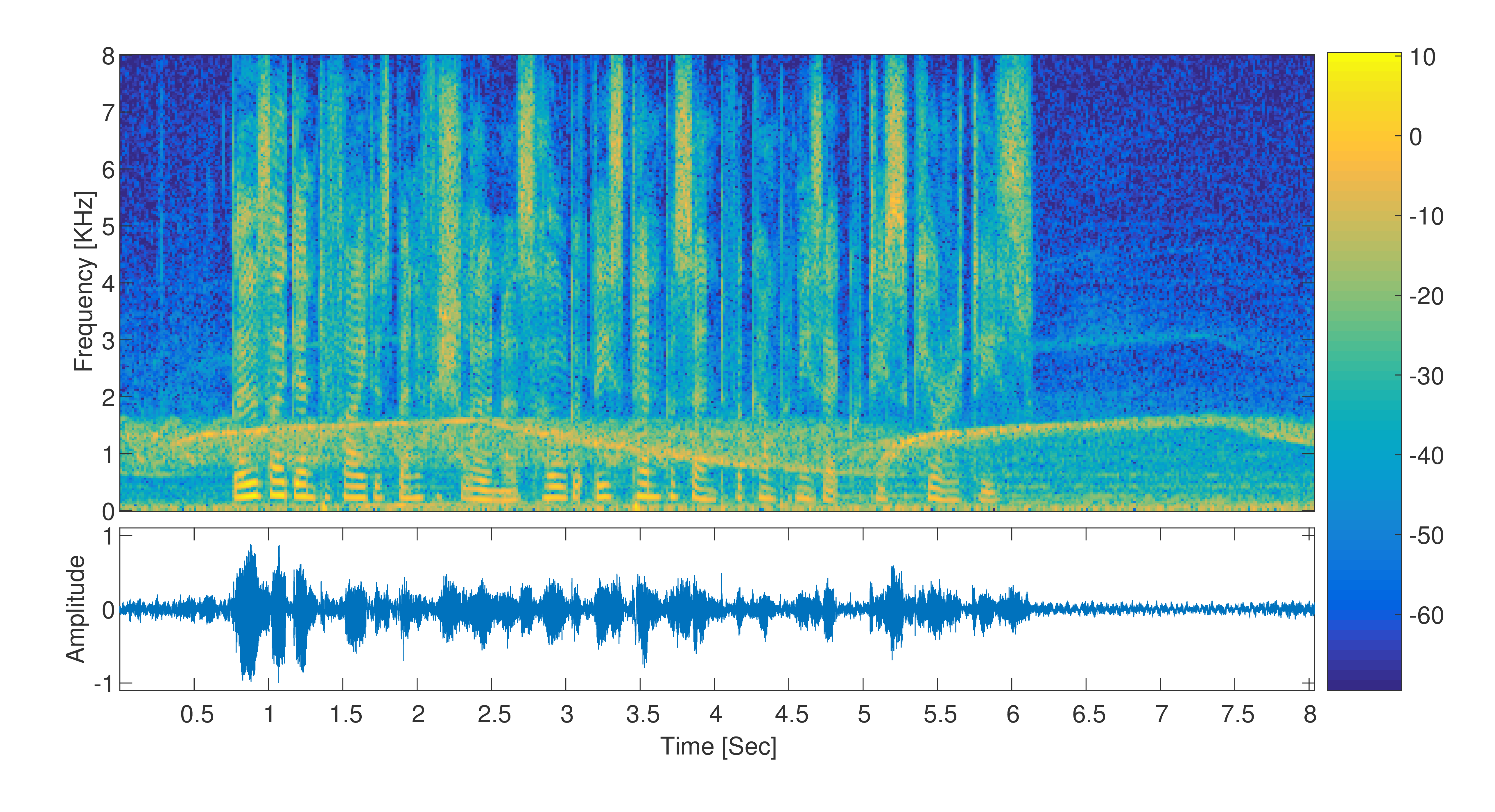}
\caption{Noisy signal (SNR=5~dB).}
\label{fig:Noisy_siren_STFT}
\end{subfigure}%
\\
\begin{subfigure}[normla]{0.5\textwidth}
\hspace{-0.62cm}
\includegraphics[width=9.4cm,height=4cm]{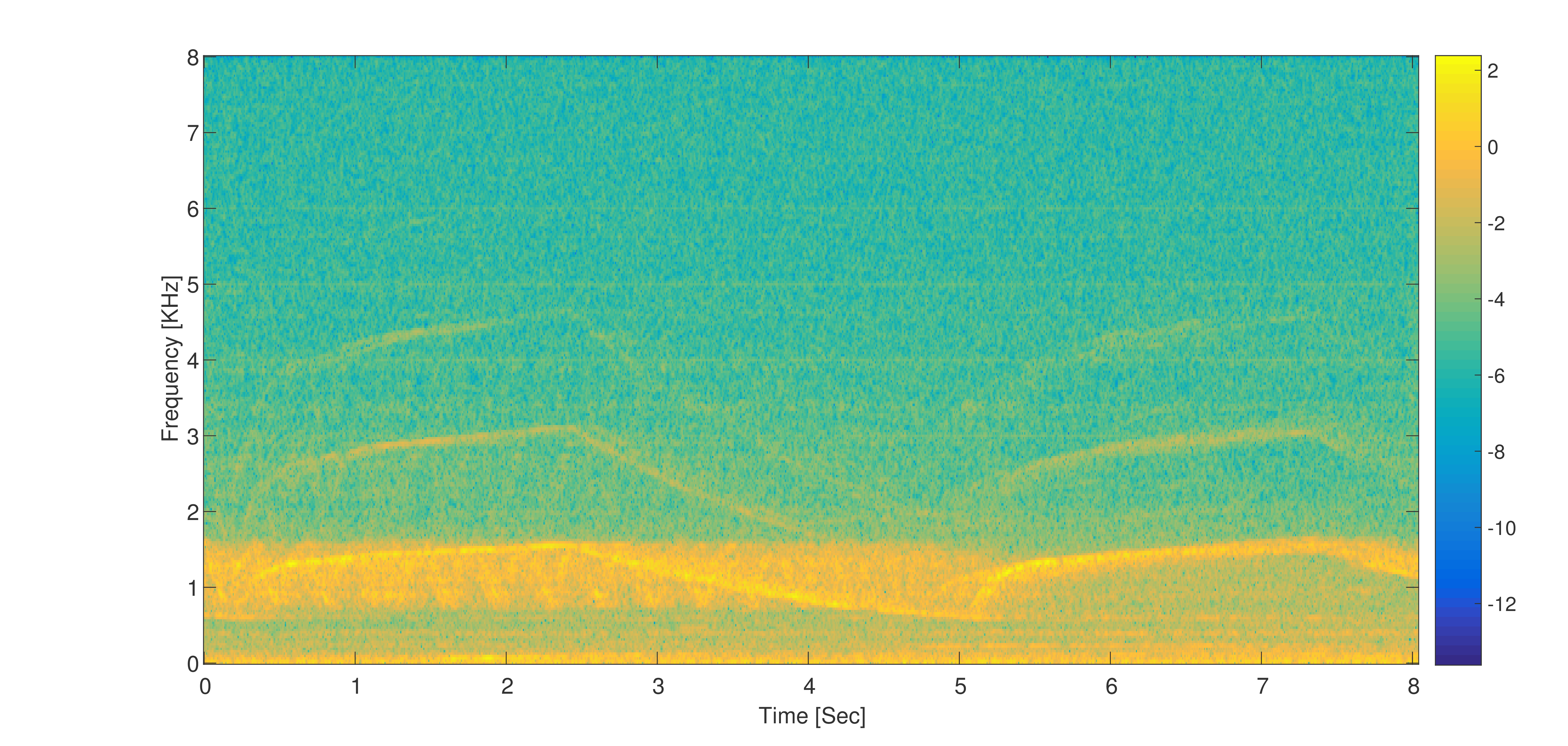}
\caption{Real noise (Log-spectrum).}
\label{fig:real_mun}
\end{subfigure}%
\begin{subfigure}[normla]{0.5\textwidth}
\hspace{-0.62cm}
\includegraphics[width=9.4cm,height=4cm,left]{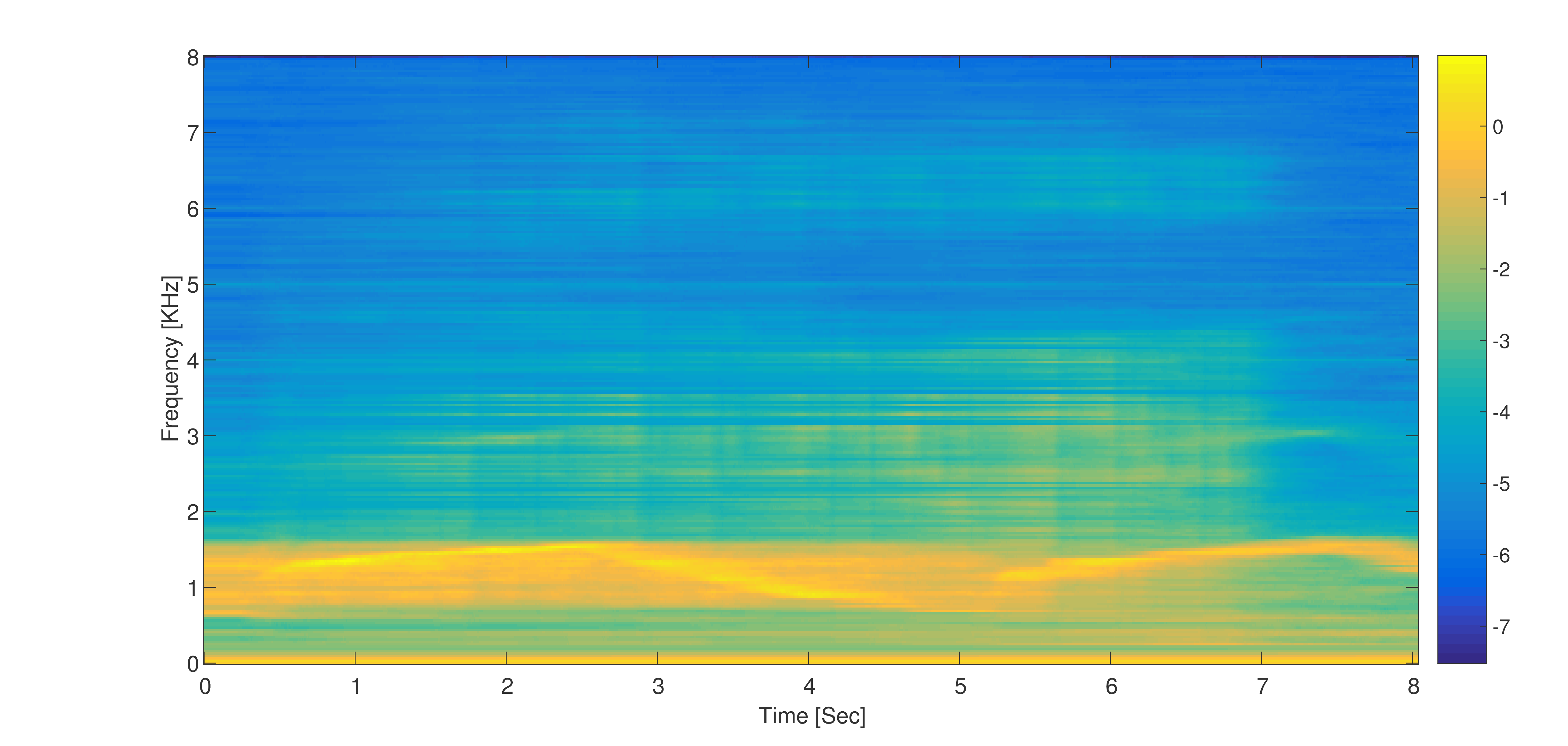}
\caption{Estimated noise (Log-spectrum).}
\label{fig:est_mun}
\end{subfigure}
\\
\begin{subfigure}[normla]{0.5\textwidth}
\centering
\includegraphics[width=\textwidth]{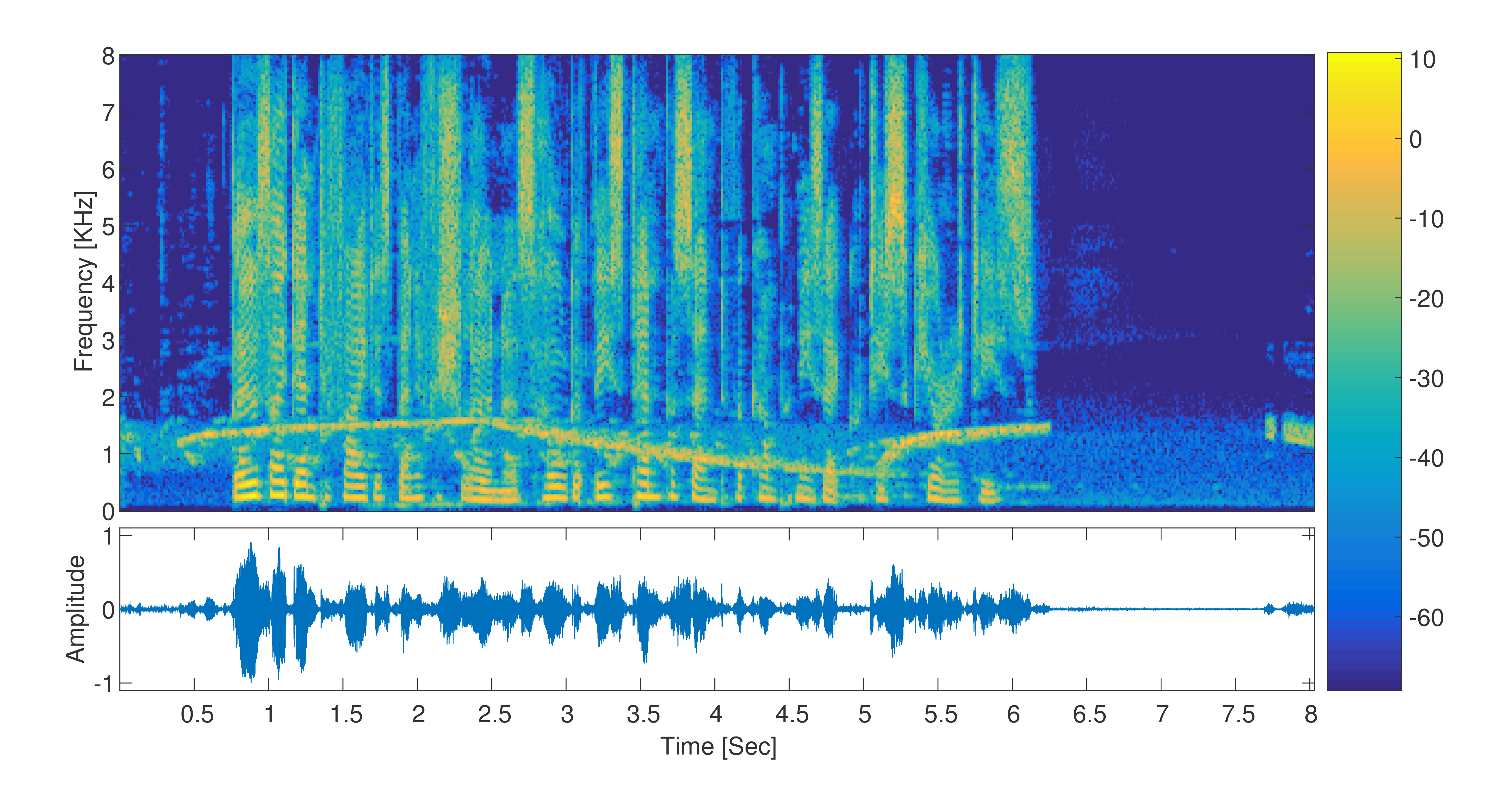}
\caption{OMLSA enhanced.}
\label{fig:OMLSA_siren}
\end{subfigure}%
\begin{subfigure}[normla]{0.5\textwidth}
\includegraphics[width=\textwidth]{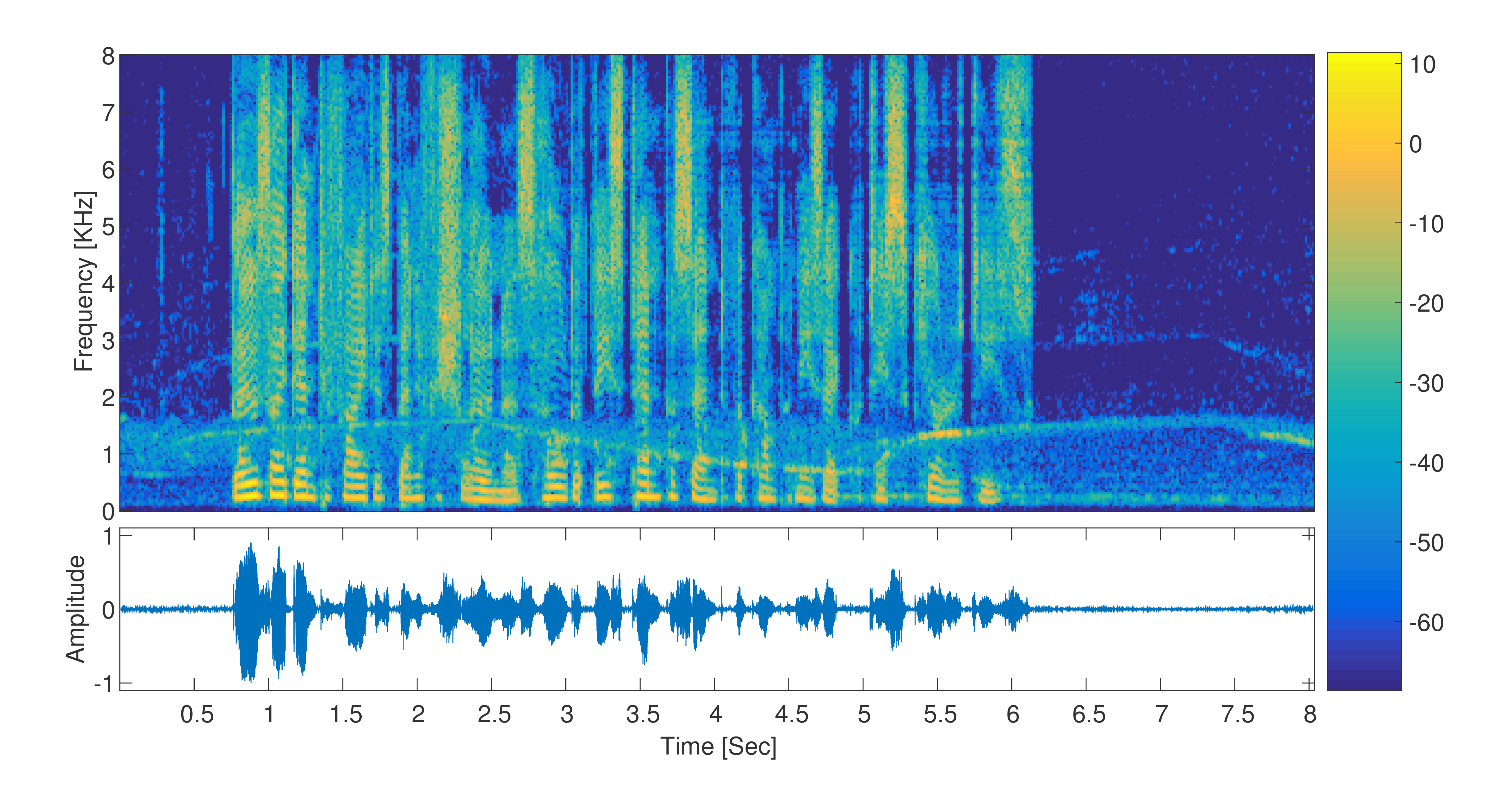}
\caption{NN-MM enhanced.}
\label{fig:ANN_MM_siren}
\end{subfigure}
\caption{Noise adaptation capabilities with highly non-stationary siren noise (SNR=5~dB), and the outputs of the \ac{OMLSA} and \ac{NN-MM} algorithms.}\label{fig:noise_adapt}
\end{figure*}

\subsection{The Noise adaptation}\label{subsec:noiseadaptresults}
In this section we examine the noise adaptation scheme described in~\eqref{noise_adapt}.  A \emph{city ambiance} noise~\cite{42} that consists of a siren and passing cars was chosen, as it is a highly non-stationary noise source with fast \ac{PSD} changes during the speech utterance. The clean and noisy signals are depicted in Figs.~\ref{fig:Clean_siren_STFT} and~\ref{fig:Noisy_siren_STFT}. The input \ac{SNR} was set to 5~dB (resulting in input \ac{PESQ}=2.124.

In Fig.~\ref{fig:real_mun} the real noise \ac{STFT} is depicted and in Fig.~\ref{fig:est_mun} its estimate using the proposed adaptation scheme and the \ac{SPP} inferred by the \ac{NN-MM} algorithm. It can be observed that the estimate is quite accurate even when the noise \ac{PSD} changes very fast. Note that during speech dominant time-frequency bins, the noise estimate cannot adapt. These adaptation capabilities are also reflected at the output of the algorithms, especially in comparison with the \ac{OMLSA} algorithm, as depicted in Figs.~\ref{fig:OMLSA_siren} and~\ref{fig:ANN_MM_siren}. We observe that the \ac{NN-MM} algorithm outperforms the \ac{OMLSA} in reducing this challenging noise. This is also indicated by the \ac{PESQ} measure. While the \ac{OMLSA} degrade the speech quality (PESQ=1.847), the proposed hybrid algorithm slightly improves it (PESQ=2.361). The reader is also referred to our website where these sound clips can be found.

\section{Conclusion} \label{sec:summer}
In this paper a novel speech enhancement scheme, denoted \ac{NN-MM}, is presented. The proposed algorithm is based on a hybrid scheme which combines phoneme-based generative model for the clean speech signal with a discriminative, \ac{NN}-based \ac{SPP} estimator.
In the proposed algorithm we try to adopt the advantages of model-based approaches and \ac{NN} approaches.
While the former usually trade-off noise reduction abilities with residual musical noise, the latter often suffer from speech distortion artifacts.

In the proposed algorithm we take advantage of the \emph{discriminative} nature of the \ac{NN} that preserves speech smoothness by using context frames. Moreover, the phoneme-based \ac{MoG} model, where each Gaussian corresponds to a specific phoneme, preserves the general phoneme structure and reduces musical noise.

The proposed algorithm requires neither noise samples nor noisy speech utterances to train. Alternatively, using the embedded \ac{NN}-based \ac{SPP}, allows for fast adaptation to fast-changing noise \ac{PSD}.

A comprehensive set of experiments demonstrate the capabilities of the proposed algorithm in both improving \ac{ASR} scores as well as objective quality measures. The \ac{NN-MM} algorithm is shown to outperform state-of-the-art algorithm (\ac{OMLSA}) for both stationary and non-stationary environmental noises and a variety of \ac{SNR} levels.

\balance


\end{document}